\def\today{\ifcase\month\or January\or
February\or March\or April\or May\or June\or
  July\or August\or September\or October\or November\or December\fi
  \space\number\day, \number\year}
\documentclass{emulateapj}

\newcommand{\TransAntiBelieve}{1.01}
\newcommand{\Flimit}{0.65}

\newcommand{\PlanetPeriodLong}{14.3053}
\newcommand{\PlanetPeriod}{14.305}
\newcommand{\PlanetRadius}{0.261} % //0.249781
\newcommand{\PlanetRadiusEarth}{2.92} % //1.62255
\newcommand{\PlanetInc}{88.8}
\newcommand{\PlanetPhi}{0.871}
\newcommand{\PlanetChiPerc}{0.893}
\newcommand{\PlanetEphemeris}{2051.4554}
\newcommand{\PlanetMassEarth}{13.6} % really 11.9
\newcommand{\PlanetMassJupiter}{0.043} % really 0.063

\newcommand{\limbone}{0.410769}
\newcommand{\limbtwo}{-0.108909}
\newcommand{\limbthree}{0.904020}
\newcommand{\limbfour}{-0.437364}

\newcommand{\NumNoiseFloor}{2}
\newcommand{\NumTopBLS}{2}
\newcommand{\Pmin}{0.5}
\newcommand{\Pmax}{14.7}
\newcommand{\NumPassedSelection}{4}

\newcommand{\EtaChoice}{0.000168}

\newcommand{\Rjupsmall}{0.20}
\newcommand{\Rjuplarge}{0.36}
\newcommand{\Rearthsmall}{2.2}
\newcommand{\Rearthlarge}{4.0}
\newcommand{\Massearthsmall}{6} % really 6.1176
\newcommand{\Massearthlarge}{35} % really 30.055586
\newcommand{\Massjupitersmall}{0.019} % really 0.019244
\newcommand{\Massjupiterlarge}{0.011} % really 0.094546

\newcommand{\MassearthlargeNeptune}{19}

\begin{document}

% \slugcomment{Draft \today}

\shorttitle{Looking for Giant Earths in the HD 209458 System}

\shortauthors{Croll et~al.}

\title{Looking for Giant Earths in the HD 209458 System:
A Search for Transits in {\it MOST}\footnotemark[1] Space-Based Photometry}

\footnotetext[1]{Based on data from the MOST satellite, a Canadian Space
Agency mission, jointly operated by Dynacon Inc., the University of Toronto
Institute of Aerospace Studies and the University of British Columbia, with
the assistance of the University of Vienna.}

\author{Bryce Croll\altaffilmark{2}, Jaymie M. Matthews\altaffilmark{2},
Jason F. Rowe\altaffilmark{2}, Rainer Kuschnig\altaffilmark{2}, Andrew
Walker\altaffilmark{3}, Brett Gladman \altaffilmark{2}, Dimitar Sasselov\altaffilmark{4},
Chris Cameron\altaffilmark{2}, Gordon A.H.
Walker\altaffilmark{5}, Douglas N.C. Lin \altaffilmark{6}, 
David B. Guenther\altaffilmark{7}, Anthony F.J. Moffat\altaffilmark{8}, Slavek M.
Rucinski\altaffilmark{9}, Werner W. Weiss\altaffilmark{10}}

\altaffiltext{2}{Dept. Physics \& Astronomy, University of British
Columbia, 6224 Agricultural Road, Vancouver, BC V6T 1Z1, Canada;
croll@astro.utoronto.ca, matthews@phas.ubc.ca, rowe@phas.ubc.ca,
kuschnig@phas.ubc.ca, gladman@phas.ubc.ca, ccameron@phas.ubc.ca}

\altaffiltext{3}{Sumus Technology Ltd.; arwalker@sumusltd.com}

\altaffiltext{4}{Harvard-Smithsonian Center for Astrophysics,
60 Garden
Street, Cambridge, MA 02138, USA; sasselov@cfa.harvard.edu}

\altaffiltext{5}{1234 Hewlett Place, Victoria, BC V8S 4P7, Canada;
gordonwa@uvic.ca}

\altaffiltext{6}{University of California Observatories, Lick Observatory, University of California, Santa Cruz, CA 95064, USA;
lin@ucolick.org}

\altaffiltext{7}{Department of Astronomy and Physics, St. Mary's
University, Halifax, NS B3H 3C3, Canada; guenther@ap.stmarys.ca}

\altaffiltext{8}{Obs du mont M\'egantic, D\'ept de physique, Univ de
Montr\'eal, C.P.\ 6128, Succ.\ Centre-Ville, Montr\'eal, QC H3C 3J7,
Canada; moffat@astro.umontreal.ca}

\altaffiltext{9}{Dept. Astronomy \& Astrophysics, David Dunlap Obs., Univ.
Toronto, P.O.~Box 360, Richmond Hill, ON L4C 4Y6, Canada;
rucinski@astro.utoronto.ca}

\altaffiltext{10}{Institut f\"ur Astronomie, Universit\"at Wien,
T\"urkenschanzstrasse 17, A--1180 Wien, Austria; weiss@astro.univie.ac.at}

\begin{abstract}

We have made a comprehensive transit search for exoplanets
down to about 2 Earth radii in the HD 209458 system, based on nearly
uninterrupted broadband optical photometry obtained with the {\it MOST}
({\it Microvariability and Oscillations of Stars}) satellite, spanning
14 days in 2004 and 44 days in 2005. We have searched these data for
limb-darkened transits at periods other than that of the known giant
planet, from about 0.5 days to 2 weeks. Monte Carlo statistical tests of
the data with synthetic transits inserted allow us to rule out additional
close-in exoplanets with sizes ranging from about \Rjupsmall \ - \Rjuplarge \ $R_{J}$
(Jupiter radii), or \Rearthsmall \ - \Rearthlarge \ $R_{\earth}$ (Earth radii) on orbits
whose planes are near
that of HD 209458b.
These null results constrain
theories that invoke lower-mass planets in orbits similar to HD 209458b
to explain its anomalously large radius, and those that predict ``hot Earths'' due to the inward migration of 
HD 209458b.
\end{abstract}


\setcounter{footnote}{0}      % otherwise corrupted numbers

\section{INTRODUCTION}
\label{SecIntro}

Since the discovery of the first exoplanet around a Sun-like star, 51
Pegasi b \citep{Mayor}, over 200 other exoplanets have been discovered
to date \citep{Schneider}. Of these, a significant number have been ``hot
Jupiters'' - gas giant planets orbiting extremely near their parent
stars. With the launch of the {\it MOST} ({\it Microvariability
\&
Oscillations of STars}) satellite (\citealt{Walker};
\citealt{Matthews})
in 2003, and the upcoming launches of the COROT (\citealt{Baglin};
\citealt{Barge})
and Kepler (\citealt{Borucki}; \citealt{Basri})
space missions, the age of transit searches with continuous space-based
photometry is now upon us. These new observatories signal that the
discovery of an Earth-sized planet orbiting a Sun-like star is a prospect
that can now be seriously anticipated.

The first transits of an exoplanet were observed in the system HD 209458,
whose planet was discovered from radial velocity measurements by Mazeh
et al. (2000).
The transits were detected by \citet{Charbonneau} and \citet{Henry}.
The discovery of this transiting planet has led to concentrated studies
of the system and hence improved knowledge of the star's characteristics
(e.g., \citealt{Wittenmyer}; \citealt{Mandel}; \citealt{Knutson}),
which is critical for the determination of the radius and other properties of the known exoplanet and any
additional transiting exoplanets found orbiting HD 209458.
%HD 209458a is an F8 dwarf with
HD 209458 is a G0 V, V=7.65 star \citep{Laughlin} with
effective temperature $T_{\rm eff} = 6000 \pm 50$ K, luminosity $L =
1.61 L_{\odot}$ \citep{Mazeh}, mass $M = 1.10 \pm 0.07 M_{\odot}$, and radius $R =
1.13 \pm 0.02 R_{\odot}$ \citep{Knutson}.

There are theoretical reasons to expect smaller exoplanets in close
orbits around HD 209458. The anomalously large size of the known
exoplanet \citep{Laughlin} has led to suggestions that the planet may
be ``puffed up" due to tidal heating involving the interaction of a
second planet (\citealt{Laughlin}; \citealt{Bodenheimer}; \citealt{BodenheimerThree}) in a resonant
orbit with HD 209458b.
\citet{Zhou} and \citet{Raymond} predict that the inward migration of hot Jupiter planets will lead to Earth-mass and 
super-Earth-mass planets
in close orbits to their parent stars in systems such as HD 209458. 
% \citet{Ida} have proposed that the dynamical stability of HD 209458b's
% close orbit to its star can be explained by other exoplanets in close
% orbits which are below the current detection threshold for radial
% velocity measurements.

Detection of exoplanets around other stars via transit measurements is
a field of increasing importance. Groundbased wide-field
photometric surveys such as TrES \citep{Alonso}, HAT \citep{Bakos},
OGLE \citep{Udalski}, and SuperWASP \citep{Street} are simultaneously
monitoring thousands of nearby bright stars for evidence of transits.
The discoveries of TrES-1 \citep{Alonso}, XO-1b \citep{McCullough}, TrES-2 \citep{ODonovan},
and HAT-P-1b \citep{BakosB}, have illustrated the feasibility
of discovering extrasolar planets via the transit method. Other
groundbased surveys have focused primarily on globular and open
clusters and have returned a series of null results
(\citealt{Weldrake};
\citealt{Hidas};
\citealt{Hood}; \citealt{vonBraun}; \citealt{Burke};
\citealt{Mochejska}). Space-based HST results returned a null-result
in the globular cluster 47 Tucanae \citep{Gilliland}, while more
recently \citet{Sahu} detected 16 transiting planet candidates near the Galactic bulge.

All groundbased transit searches are inherently limited by time
sampling: the day-night cycle, weather and sometimes observing time allocation.
In addition, their photometric precision means they are sensitive to only
planets similar in scale to our giant planets. Space-based transit searches from satellites
in inclined equatorial orbits (such as HST) also have limited time
sampling. However, space-based transit surveys from platforms with
large continuous viewing zones (CVZ) on the sky offer the potent combination
of excellent extended-time coverage and high photometric precision.

In this paper, we describe such a search for transits in the HD 209458
system with the {\it MOST} satellite, with nearly continuous time
coverage over two epochs spanning two and six weeks in 2004 and 2005,
respectively. The methods refined and developed here to take advantage
of this precise, high duty-cycle photometry allow us to reliably
detect transits as shallow as $\sim$$0.5$ mmag, corresponding to an
exoplanet as small as about 0.2 $R_{J}$, or 2 $R_{\earth}$. These
methods are also applicable to other space-based transit searches,
such as future searches with {\it MOST}, as well as those planned with
the COROT and Kepler space missions.

In $\S$\ref{SecMOST} the {\it MOST} photometry of HD 209458 is
described. The transit search technique is discussed in
$\S$\ref{SecTrans}. The Monte Carlo statistics used to estimate the
sensitivity of the transit search are specified in $\S$\ref{SecMonte}.
The transit search routine is applied to the {\it MOST} HD 209458 data
set in $\S$\ref{SecHDTrans}, and the results are presented in
$\S$\ref{SecDiscuss}, including a discussion of the impact of these
results on theories relating to other putative exoplanets in the system, and 
the characteristics of the
known exoplanet HD 209458b.

\section{{\it MOST} Photometry of HD 209458}
\label{SecMOST}

The {\it MOST} satellite was launched on 30 June 2003 and its initial
mission is described by \citet{Walker} and \citet{Matthews}. A
15/17.3-cm Rumak-Maksutov telescope feeds two CCDs, one originally
dedicated to tracking and the other to science, through a single,
custom, broadband filter (350 -- 700 nm). {\it MOST} was placed in
an 820-km circular Sun-synchronous polar orbit with a period of
101.413 minutes. From this vantage point, {\it MOST} can monitor
stars in a continuous viewing zone (covering a declination range
$+36^{\circ} \leq \delta \leq -18^{\circ}$) within which stars can
be monitored without interruption for up to 8 weeks. Photometry of
very bright stars ($V \leq 6$) is obtained in Fabry Imaging mode, in
which a Fabry microlens projects an extended image of the telescope
pupil illuminated by the target starlight to achieve the highest
precision (see Matthews et al. 2004). Fainter stars (down to about
$V \sim 12$) can be observed in Direct Imaging mode, in which 
defocused images of stars are monitored in Science CCD subrasters
(see \citealt{Rowe}).

{\it MOST} observed HD 209458 in Direct Imaging mode over two separate epochs: for 14
days during 14 - 30 August 2004 \citep{Rowe}, and 44 days during 2
August - 15 September 2005 \citep{Rowe2}. The first run was a trial,
and testing of new onboard software led to several data interruptions,
reducing the overall duty cycle of the raw photometry to about 85\%.
The second run returned data without significant interruptions,
achieving an overall duty cycle of the raw photometry of 97\%.
The exposure time of the observations was 1.52 sec, and the sampling
rate was 10.0 sec for both epochs of data. Approximately 112,000 and
361,000 individual observations were taken in the 2004 and 2005
epochs, respectively.

The reduction of the raw photometry downloaded from the satellite
(and converted into FITS files) for both epochs was performed by JFR.
The reduction is similar to that applied to groundbased CCD
photometry, but is nondifferential, and incorporates both aperture
and PSF (point spread function) fitting. It corrects for cosmic ray
hits (especially frequent during satellite passages through the South
Atlantic Anomaly [SAA]), the varying background due to scattered
earthshine modulated at the satellite orbital period, and
flatfielding effects. Those data reduction techniques used are fully
described in \citet{Rowe} and \citet{Rowe2}.

\subsection{Additional filtering and selection of the data}
\label{SecMOSTReduce}

For this transit search, additional filtering and selection of the photometry was
done to optimize the data for the application of Monte Carlo statistics.
The filtering is an automated part of the transit search routine, and is thus briefly summarized
in $\S$\ref{SecTransReduce}. This filtering step is described in detail here.

Data obtained during passages through the SAA are conservatively
excised from the light curve, due to the increased photometric scatter
during those MOST orbital phases, without reducing seriously the phase
coverage of the exoplanetary periods searched. The transits of
the known giant planet HD 209458b are removed, at the orbital period $P$ = 3.52474859d determined by \citet{Knutson}.
Because of the modulation of stray earthshine
with {\it MOST}'s 101.4-min orbital period, the data were phased to that
period, and segments showing the most noticeable effects from stray
light were also removed. The coverage in phase
per 101.4-min {\it MOST} orbit following this cut 
was 63\% and 58\% for the 2004 and 2005 epochs,
respectively.
The stray light background can also be modulated at a period of 1 day and its first harmonic (due to the 
Sun-synchronous nature of the MOST satellite orbit), so 
sinusoidal fits with periods within 1\% of 1 and 0.5 d were subtracted from the data.
After these cuts, any remaining outliers
greater than $6\sigma$ were excised. 
This sigma cut removed very few data-points, as few points were such extreme outliers.
The magnitude of the injected transits ($\S$\ref{SecMonte}) was always
at a level much less than this sigma-cut.
The data were also
median-subtracted, for reasons outlined in $\S$\ref{SecTransDetect}.
The automatic filtering and selection step removed 45\% of the original data (mostly due to the
SAA and stray light corrections). 
The root mean square of the data following this additional filtering and selection step is 0.0035 mag.

\begin{figure}
\includegraphics[width=2.1in, height = 7.0in, angle=270]{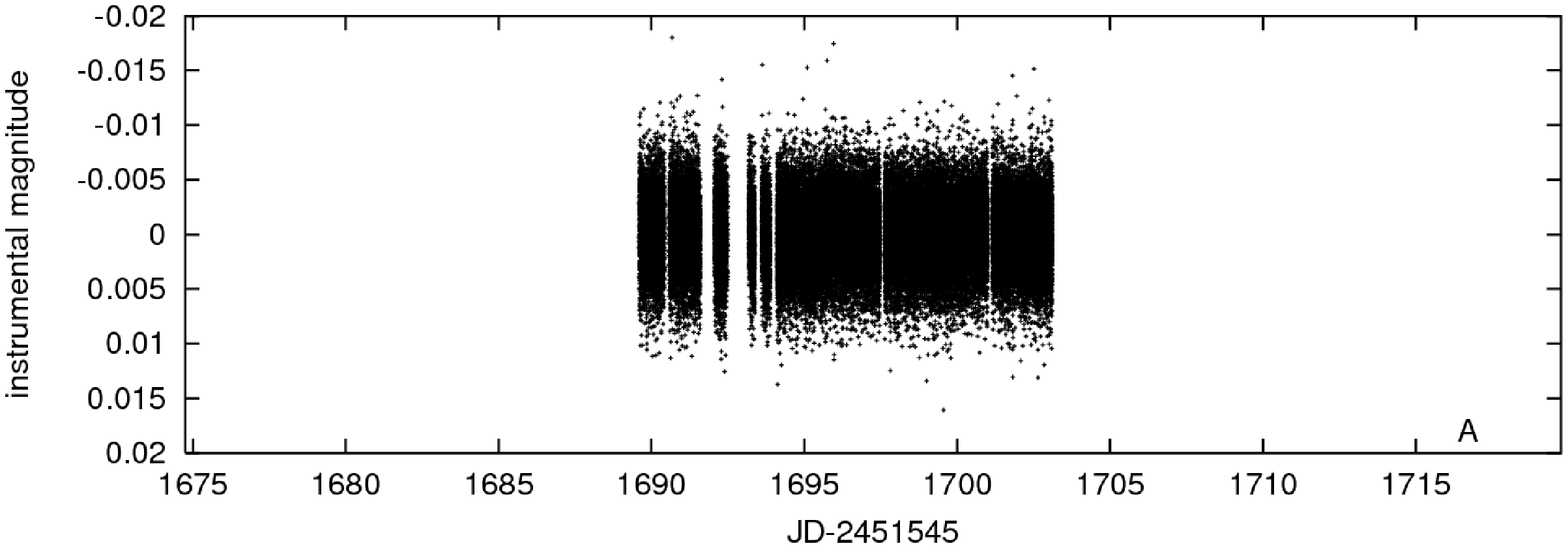}
\includegraphics[width=2.1in, height = 7.0in, angle=270]{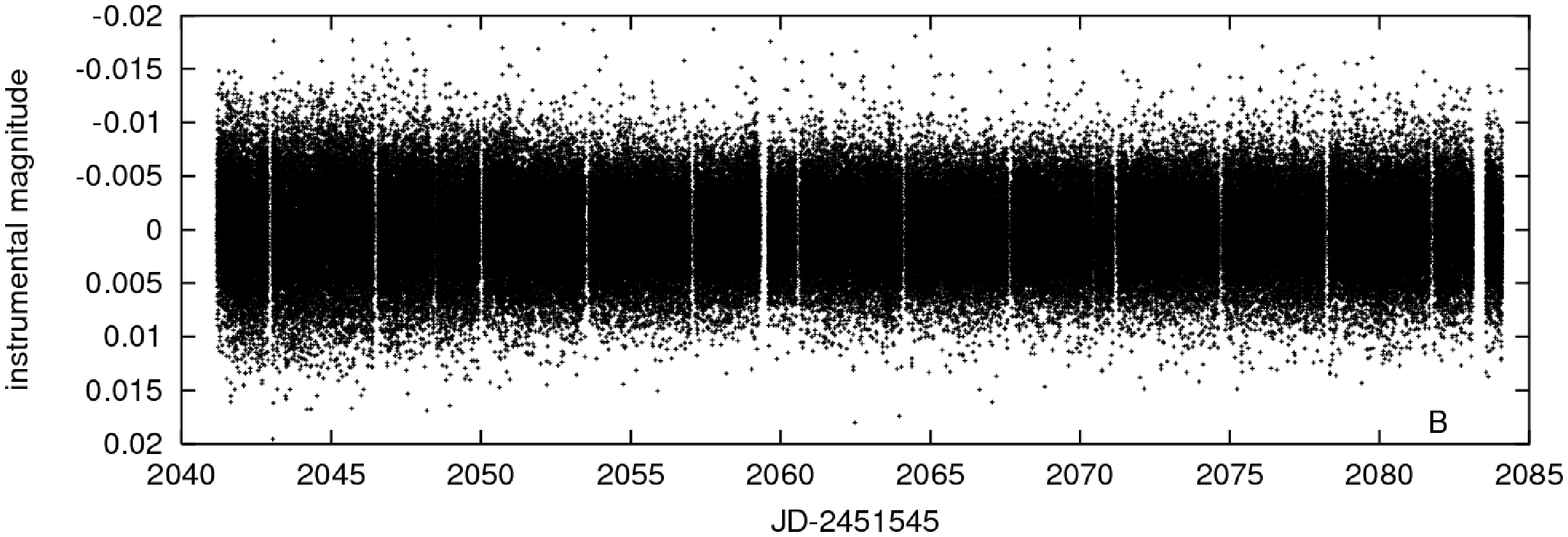}
\includegraphics[width=2.1in, height = 7.0in, angle=270]{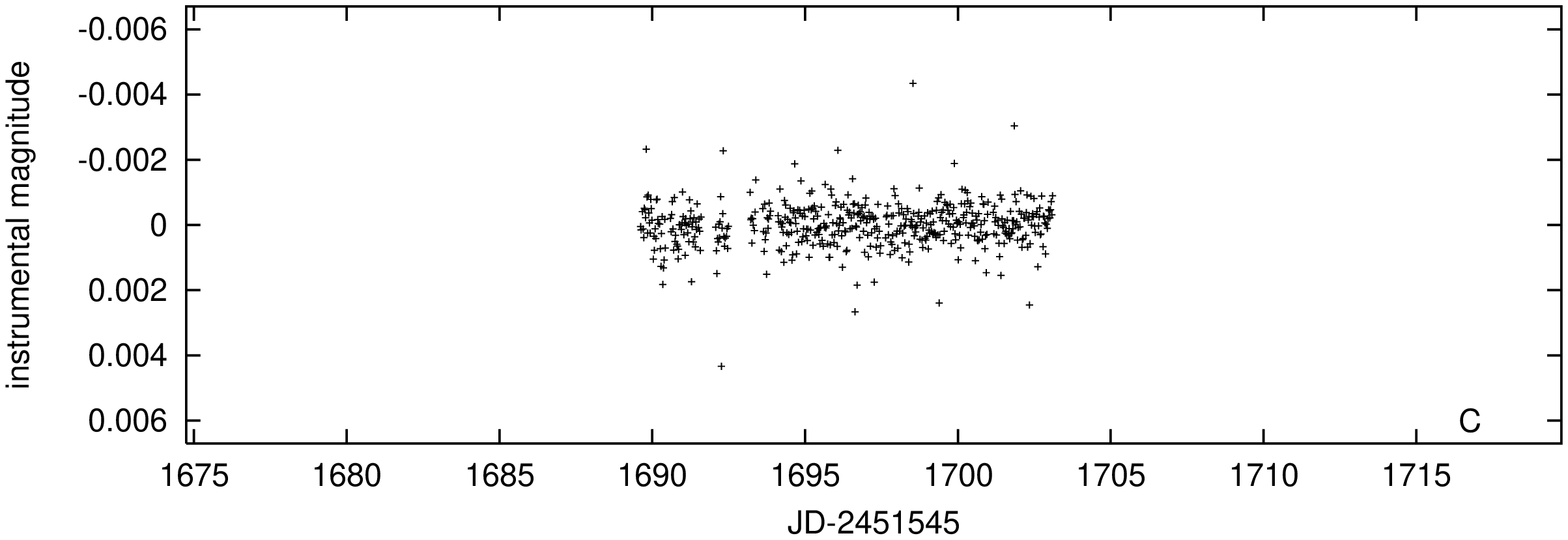}
\includegraphics[width=2.1in, height = 7.0in, angle=270]{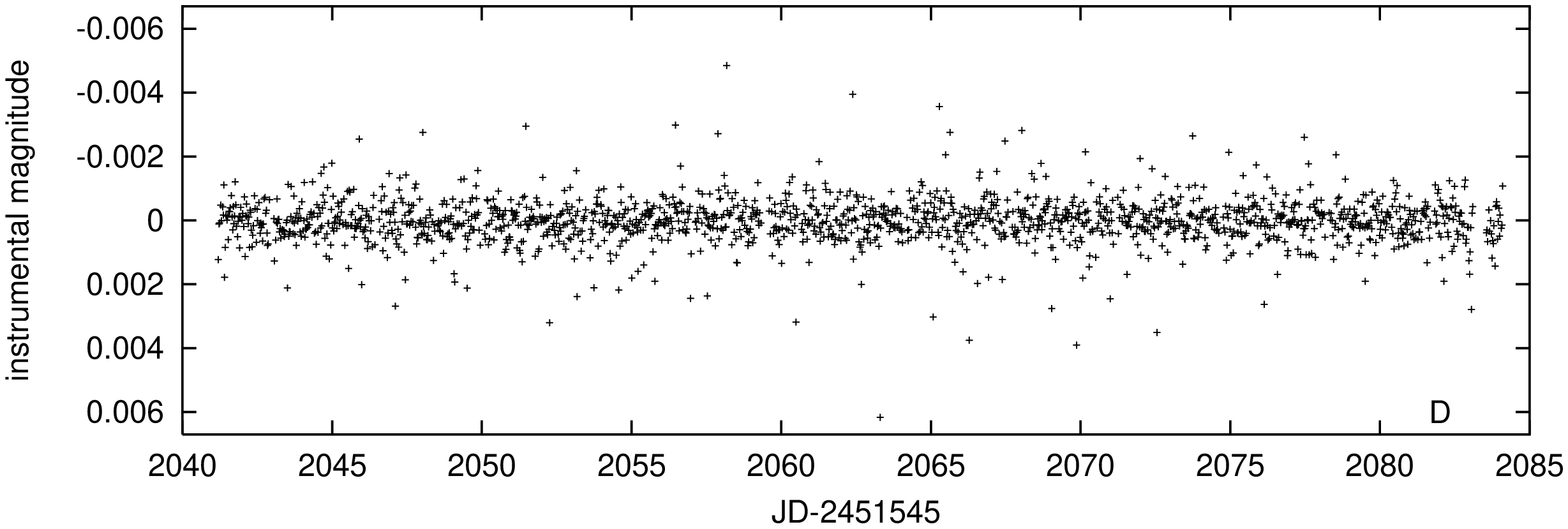}
\caption{
	{The 2004 (A) and 2005 (B) HD 209458 {\it MOST} data-sets following the initial reduction and the automatic
	filtering step. The same data are shown binned in 30 minute intervals at a different vertical scale for 2004 (C) and 2005 (D). 
	}  
	\label{FigData}
}
\end{figure}

The resulting HD 209458 light curves are plotted unbinned, and binned in 30-min intervals in Figure
\ref{FigData}. These filtered data (and the original reduced data) can
be downloaded from the {\it MOST} Public Data Archive at
www.astro.ubc.ca/MOST.

\section{Transit Search Algorithm}
\label{SecTrans}

Our transit search routine has been developed and refined to take
specific advantage of the unique time coverage of the {\it MOST} data.
For HD 209458 in particular, it is designed to make use of the well
determined stellar and exoplanetary orbital parameters of the system.
The search is intended to run automatically without user intervention
to facilitate the generation of Monte Carlo statistics of the detection
thresholds and their corresponding significances.

The entire routine (described in the following subsections) includes:
($\S$\ref{SecTransReduce}) an automatic filtering and selection step; ($\S$\ref{SecTransBLS}) a transit
search algorithm, adapted from the EEBLS (Edge Effect Box-fitting Least
Squares) algorithm of \citet{KovacsWEB} sensitive to box-shaped
transits; ($\S$\ref{SecTransSelect}) selection criteria, to narrow the EEBLS candidates
to a select few that can be tested for astrophysical plausibility;
($\S$\ref{SecTransFit}) transit model fitting,
to identify realistic limb-darkened transit candidates for exoplanets
of various radii, orbital periods and inclination angles;
and finally
($\S$\ref{SecTransDetect}) detection criteria, to differentiate between likely transiting
candidates and false positives.
This automated search routine can also
be applied to other transit search experiments in which the characteristics
of the target stars are well known or inferred.

\subsection{Automated filtering and selection scheme}
\label{SecTransReduce}

The initial reduction scheme of JFR as described in $\S$\ref{SecMOST}
is almost entirely independent of the magnitude of an individual
data-point. Instead other indicators such as the sky background level
are used to remove or adjust the data, or the associated uncertainty
of an individual data-point.
Therefore, inserting synthetic transits in the light curve after
this initial reduction to produce Monte Carlo statistics is a valid
procedure. To ensure proper application of those Monte Carlo
statistics for this and other data sets, we incorporate an automated
data filtering and selection scheme in the transit search to filter out stray earthshine, and to 
remove additional
outliers and the dominant transits of the known exoplanet. This process was
described in detail in $\S$\ref{SecMOSTReduce}. It is briefly summarized here to ensure that it is
obvious that this process
is an automated part of the transit search routine applied to the Monte Carlo statistics to come.
The thresholds and amplitudes
used in this automatic filtering scheme have been optimized for their application to the 
{\it MOST} HD 209458 data; obviously these thresholds and amplitudes would have to be similarly optimized 
for their application
to other {\it MOST} data-sets. 

\subsection{EEBLS (Edge Effect Box-fitting Least Squares) algorithm}
\label{SecTransBLS}

To search for transits in the HD 209458 system at periods other than
that of the known exoplanet HD 209458b, we use a slight modification
of the Box-fitting Least Squares (BLS) algorithm \citep{Kovacs}, namely the Edge Effect
Box-fitting Least Squares (EEBLS) algorithm \citep{KovacsWEB}. The
EEBLS routine flags putative transit candidates by
searching for box-shaped transits in a light curve. The signal strength
of a putative transit is indicated by the Signal Residue (SR), as
defined in \citet{Kovacs}. The SR is roughly equivalent to the $\chi^{2}$ for a box-shaped transit as indicated in the appendix of \citet{Burke}.
The EEBLS routine returns the period, phase, and drop in magnitude of
the supposed transit. The EEBLS algorithm was chosen because it has been used extensively in similar searches
(e.g. \citealt{Burke}; \citealt{Mochejska}; \citealt{Sahu}),
and quantitative comparisons indicate it is as good as others in the 
literature \citep{Tingley}.

As is the case with all computational routines that must handle a
large number of data and an extensive search parameter grid, sufficient
resolution must be balanced against computational duration. To
decrease the latter, we employ even logarithmic period spacing of
the search grid as suggested by \citet{Burke}. With even logarithmic
period spacing, the subsequent period searched, $P_{2}$, is related to
the previous, $P_{1}$, by $P_{2}$ = $P_{1}$$(1+\eta)$, where $\eta$
$\ll$ 1. Although even frequency spacing was used in the original
formulation of BLS by \citet{Kovacs}, even logarithmic period spacing
retains high sensitivity to transits with both short and long orbital periods without
increasing computational duration. Both even frequency and even
logarithmic period spacing are valid with sufficient resolution.
We chose the value $\eta$ $=$ \EtaChoice \ to attain the especially fine period resolution needed to 
resolve putative transits occuring in the two epochs of {\it MOST} data separated by the near one-year gap.

In this work, we search the data set for periods greater than $P_{min}
= \Pmin d$, and less than $P_{max} = \Pmax d$. The maximum period limit
is set by our requirement to observe at least 3 hypothetical transits
of a putative planet in the longer 2005 run (44 days). The minimum
period was set to be 0.5$d$, likely below the astrophysically reasonable
period of the orbit of a stable exoplanet in the HD 209458 system, as it would correspond to a semi-major 
axis less then $\sim$2.5 times the stellar radius.

The EEBLS routine was also modified to enhance its detection
sensitivity. For each trial period, the EEBLS routine was changed to
record the most significant box-car dimming (transit) and box-car
brightening signal independently. The original formulation recorded the
most significant box-car event, regardless of whether it was a dimming
or a brightening. This resulted in a modest decrease in sensitivity as 
the noise floor increased because the algorithm was automatically
including brightening events that could not possibly be transits.
This reformulation allows one to better differentiate possible transits,
from systematic brightening events. 

In addition the minimum and maximum fractional transit lengths, $Qmi$ and $Qma$ respectively, that
the EEBLS algorithm searches each trial period for were set to be variable, rather than constant as in the original formulation.
The small planet approximation of \citet{Mandel}, as discussed below, was used to determine the approximate
maximum fractional transit length for each trial period, $Qm_{P}$.
For this approximation a planet with radius $R_{p}$ = $0.4$$R_{J}$, and with inclination angle $i$ = 90$^{o}$ was used to determine this maximum fractional transit length.
This change is significant because the difference in the maximum fractional transit length falls from nearly
$Qm_{P}$ $\approx$ 0.13 at the minimum period of 0.5$d$, 
to $Qm_{P}$ $\approx$ 0.015 at the maximum period of 14.7$d$.
This change allows one to thus use the known characteristics of the star that is investigated to precisely tailor
the EEBLS search routine. 
$Qmi$ and $Qma$ were set significantly lower, and slightly larger, respectively,
than this maximum fractional transit length for each period; 
that is $Qmi$ was set as 0.75 times as large, and
$Qma$ was set as 1.1 times as large as the maximum fractional transit length for each period.
These values of 0.75 and 1.1 times
as large as the maximum fractional transit length were chosen
because they span the parameter range of interest from a 
large edge-on planet to a small grazing planet.   

The final change that was implemented was to set a variable number
of bins in the folded time series at each test period, $Nb$, rather than the constant number as in the original formulation. This was again related to the
fact that the maximum fractional transit length varies significantly from the minimum to maximum period investigated in this formulation.
The decision was
made to ensure that there was a constant number of bins in the folded time series during the in-transit component for the smallest fractional transit length, $Qmi$,
for each trial period. 
This constant number of bins was set arbitrarily at 20,
as it was believed this was large enough to properly determine the phase of a putative transiting planet.
That is for each trial period $Nb$ was set as 20.0/$Qmi$. $Nb$ thus varied from as low as $\approx$200
for the minimum period of 0.5$d$, to as high
as $\approx$1570 for the maximum period of 14.7$d$.
These changes not only increase the sensitivity of the algorithm to long-period events, but also marginally decrease the computational duration
of the EEBLS algorithm. The EEBLS input parameters are summarized in Table \ref{TableBls}.
The modified version of this algorithm in C is available at:
www.astro.ubc.ca/MOST/EEBLSmodified.cpp

\begin{deluxetable}{ccc}
\tabletypesize{\footnotesize}
\tablecaption{EEBLS input parameters \label{TableBls}}
\tablewidth{0pt}
\tablehead{\colhead{var}&	\colhead{definition} &	\colhead{value} \\}
\startdata
$Np$ 		& Number of Period Points searched 					& 20 000 			\\ 
$\eta$ 		& logarithmic period step 						& \EtaChoice			\\ 
$P_{min}$	& min. period submitted to EEBLS				& \Pmin$d$ 			\\	
$P_{max}$	& max. period submitted to EEBLS				& \Pmax$d$			\\
$Qmi$		& min. fractional transit length 			& $0.75$$\times$$Qm_{P}$ 	\\
$Qma$ 		& max. fractional transit length 			& $1.1$$\times$$Qm_{P}$		\\
$Nb$ 		& Number of phase bins			 		& 20.0/$Qmi$			\\
\enddata
\end{deluxetable}

\subsection{Transit Selection Criteria}
\label{SecTransSelect}

The EEBLS algorithm returns the Signal Residue (SR) versus period for each of the $Np$ periods investigated.
Selection criteria have been developed to pick
the most likely
genuine transit candidates. These selection criteria are described here.
Of all the periods, only those
that are strong local maxima in SR versus period are flagged for
further investigation. A strong local maximum is defined as one
for which SR is greater than the 0.003
$\times$ $nP$ neighbouring period points in both directions. 
Of the strong local maxima, the candidates with the
highest \NumTopBLS \ SR signals are chosen for further analysis.
As we often noted that simply picking the top
SR values biases one's selection against short-period planets of
small radius, a second selection criterion was implemented 
to ensure adequate sensitivity to short-period events ($\sim$0.5 $<$
P $< \sim$4.0). The top \NumNoiseFloor \ candidates with SR values
$1.8 \times$ greater than the SR noise floor were also selected
for additional study. The SR noise floor is defined as the mean of
the SR values for all period points within 0.2 $\times$ $nP$ of the
period of interest. These two selection criteria narrowed the list
of prospective candidates to a maximum of four of the most likely transit
candidates. Extensive statistical tests (described in
$\S$\ref{SecMonte}) both verified and refined the above criteria.

\begin{figure}
\includegraphics[height = 6in, width = 4in, angle=270]{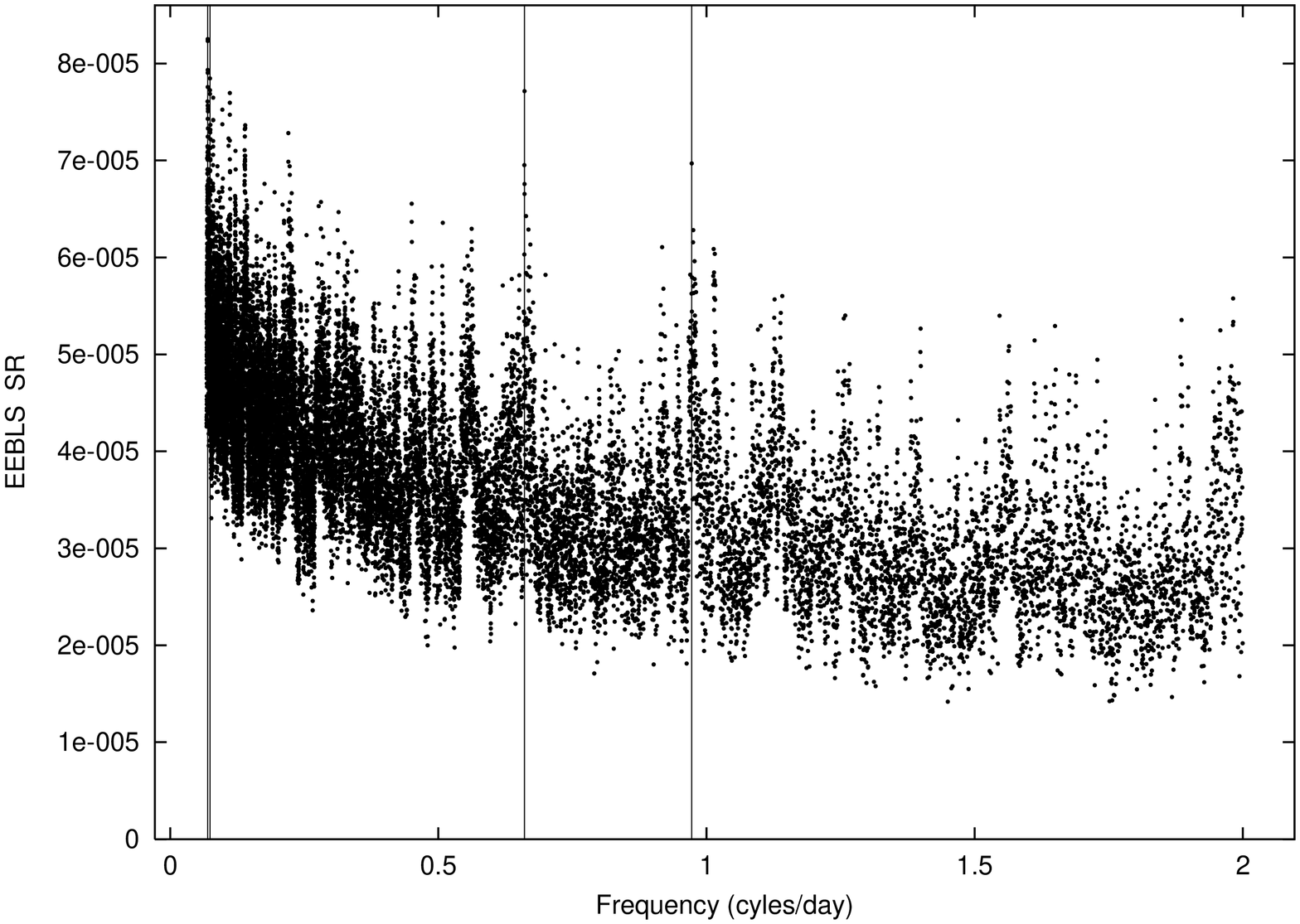}
\includegraphics[height = 6in, width = 4in, angle=270]{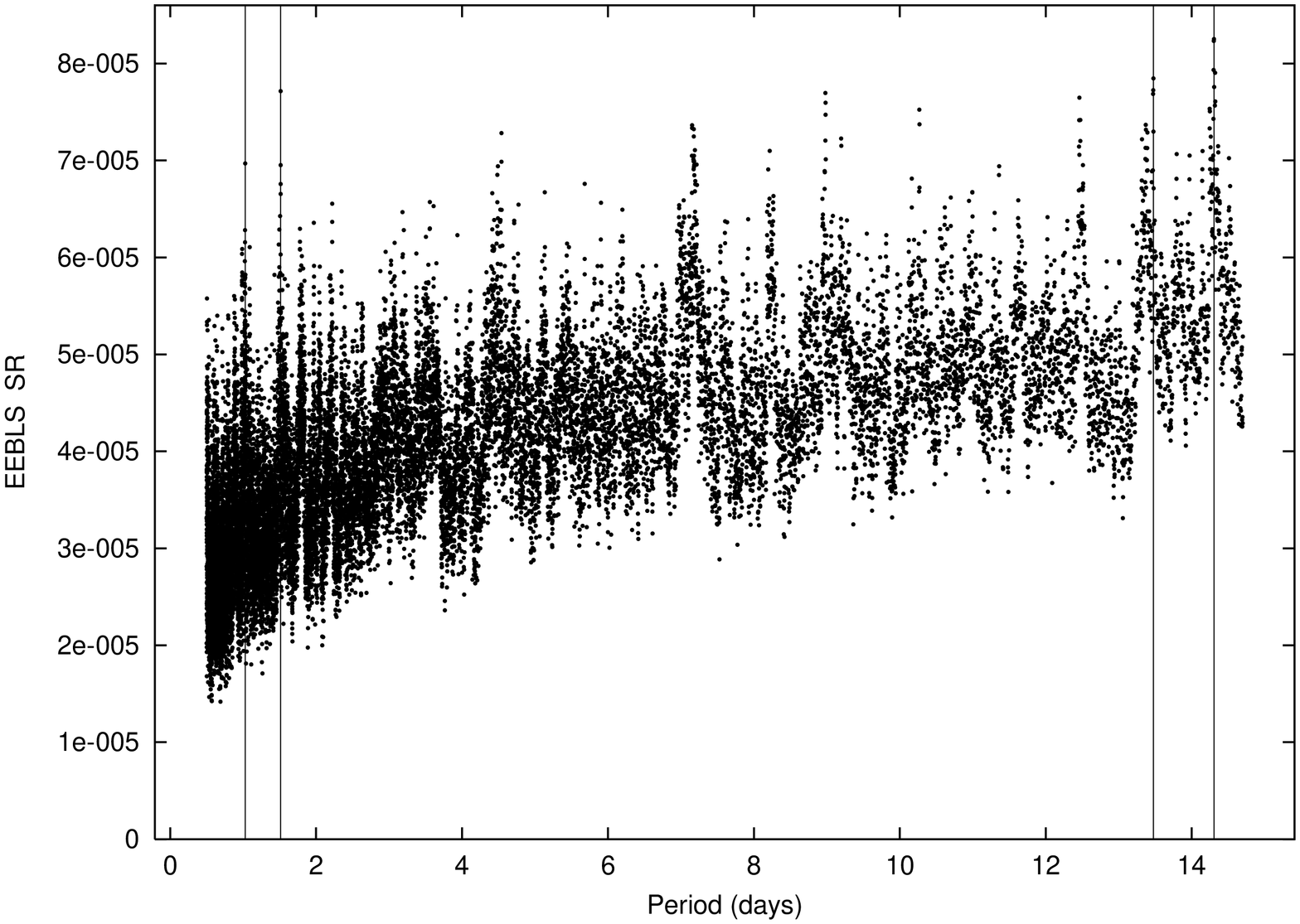}
\caption{{EEBLS spectrum of the total {\it MOST} HD 209458 data-set, plotted in frequency (top),
and period (bottom).
The \NumPassedSelection \ candidates that passed the transit selection
criteria as discussed in $\S$\ref{SecTransSelect} are shown by the solid vertical lines.
In the frequency plot
the lines marking the two candidates with the lowest frequencies appear as one.}  
\label{FigTransSelect}}
\end{figure}

The EEBLS SR spectrum for the {\it MOST} HD 209458 2004 and 2005 light
curves
is shown in Figure \ref{FigTransSelect}; the candidates that passed the selection
criteria for these light curves are also identified in Figure \ref{FigTransSelect}. 

We also optimize the determination of the exact period, $P$, and phase, $\phi$, for these candidates by 
rerunning the EEBLS algorithm with greater period, and then phase resolution,
in the immediate period space
of the candidate event.
First, we run the EEBLS algorithm with greater period resolution from a minimum to maximum period
of 4 period points less to 4 period points greater, respectively, than the best-fit period.
An order of magnitude more period points are then used 
(100 points), to more accurately determine the period, at which the greatest SR signal is returned. 
The period that returns the highest EEBLS SR is then investigated with the EEBLS algorithm
at this one specific period
with 5 times as many phase points as previously.
These two steps are of negligible computational duration, but considerably increase the accuracy
with which the period, and phase are returned.  

\subsection{Realistic Transit Fitting}
\label{SecTransFit}

Since the discovery of the giant exoplanet HD 209458b, there has been
intensive study of the star HD 209458, such that
its basic characteristics
are now well known (\citealt{Wittenmyer}; \citealt{Knutson}).
This allows for considerable improvement in the
shape, duration, and depth of the transit model
as opposed to the original box-shaped transit  
used in the EEBLS routine.
Specifically, we search for realistic
limb-darkened transits of a specific period, $P$, planet radius,
$R_{p}$, orbital inclination angle, $i$, and phase of transit, $\phi$.
Our model uses the small planet approximation of \citet{Mandel} to set
the shape, length and depth of the transit. The small planet approximation of \citet{Mandel} is valid for $\frac{R_{P}}{R_{*}} < 0.1$, and
thus should be valid for HD 209458 for all
transiting planets below $R_{P}$ $\approx$ 1.1 $R_{J}$. 
We have only searched for
transiting planets whose orbits are circular ($e = 0$).; The impact of eccentricity is discussed in $\S$\ref{SecMonte}. The HD 209458
system parameters used for our analysis are given in Table
\ref{TableFit}.

\begin{deluxetable}{ccc}
\tabletypesize{\footnotesize}
\tablecaption{System parameters \label{TableFit}}
\tablewidth{0pt}
\tablehead{\colhead{var}&	\colhead{definition} &	\colhead{value} \\}
\startdata
$R_{*}$ & Stellar radius & 1.125 $R_{\odot}$ \tablenotemark{a} \\
$M_{*}$ & Stellar Mass & 1.101 $M_{\odot}$ \tablenotemark{a}\\
$c_{1}$ & non-linear limb-darkening parameter 1 &  \limbone \tablenotemark{b} \\
$c_{2}$ & non-linear limb-darkening parameter 2 &  \limbtwo \tablenotemark{b}\\
$c_{3}$ & non-linear limb-darkening parameter 3 &  \limbthree \tablenotemark{b}\\
$c_{4}$ & non-linear limb-darkening parameter 4 &  \limbfour \tablenotemark{b}\\
\enddata

\tablenotetext{a}{Parameters obtained from \citet{Knutson}}
\tablenotetext{b}{These limb-darkening parameters are fully described in \citet{Mandel} and obtained from \citet{Knutson}
and \citet{Rowe2}}

\end{deluxetable}

The transit candidates are
then fitted by our limb-darkened transit model. For the period $P$
returned by the EEBLS algorithm, two inclination angles $i$ are
investigated. One inclination angle is edge-on ($i = 90^{\circ}$), while the other is 
evenly spaced between the minimum inclination that would produce a
transit (cos$i$ $\sim$ $R_{*}/a$, where $a$ is the semi-major axis),
and an edge-on ($i = 90^{\circ}$) orbit. For short period events, where the minimum
inclination that would produce a transit is below 78$^{o}$, the minimum inclination 
for the initial guess
is arbitrarily set to 78$^{o}$, as it was found this increased the detection efficiency.
For example a period of approximately 4 days will have a minimum inclination angle of 84$^{o}$, and thus the
two guesses will be 87$^{o}$, and 90$^{o}$, respectively. 
For these inclination angles,
the planet radius $R_{p}$, phase $\phi$, and inclination $i$, of a putative transit are fit through a
Marquardt-Levenberg non-linear least squares algorithm
(\citealt{Marq}; \citealt{Leven}; \citealt{Numer}). Different
starting values for the orbital inclination are used, since it was
found through Monte Carlo analysis of simulated transits ($\S$\ref{SecMonte})
that this improved the efficiency in returning the correct value of
$i$. The initial guesses in radius, $R_{p}$, and phase, $\phi$, given to the Marquardt-Levenberg algorithm
are the characteristics returned by the EEBLS routine, where the dimming in
brightness is translated into the size of a hypothetical edge-on
planet using the known stellar radius. Limits were placed on the
range of parameters that the Marquardt-Levenberg algorithm could
explore, away from the starting points returned by the EEBLS
algorithm, to ensure that the fit would remain in the parameter space that was
flagged as significant by the EEBLS algorithm.
These were $\Delta R_{p} \approx 0.10$, $\arccos(R_{*}/a)
\leq i \leq 90.0^{o} $, and $\Delta \phi \approx 0.05$.
Fitting for the period as well using the Marquardt-Levenberg non-linear
least squares algorithm was not attempted here, as the period optimization
discussed above was deemed sufficient.
The values of $R_{p}$, $i$ and $\phi$
that produce the minimum $\chi^{2}$ value are recorded as the best
transit candidate for that period.

An attempt was made to bypass the EEBLS algorithm altogether and
solely use the above fitting method to search for realistic
limb-darkened transits in the {\it MOST} HD 209458 data. However, this
approach proved very computationally intensive, and so we adopted
and recommend the compromise described above, which achieves sufficient accuracy without
significant increase in computational duration.

\subsection{Transit Detection Criteria}
\label{SecTransDetect}

We have developed quantitative criteria to differentiate bona-fide
transiting planet candidates from false positive detections, without
the need for detailed visual inspections of every set of candidate
events in the light curve. We compare each modeled transit
candidate to an anti-transit (brightening) model, and a constant
brightness model. The latter assumes a star of constant magnitude,
given by the mean of the light curve. The transit model consists of
an out-of-transit component and an in-transit component (modeled as described
in $\S$\ref{SecTransFit}). The out-of-transit component has a
constant brightness level given by the median of the light curve; 
the median was chosen, as opposed to the mean, because it should remain relatively unchanged
regardless of whether one or more other transiting planets are
present in the HD 209458 system. 
During transit a realistic limb-darkened transit of a spherical planet is used with parameters 
$R_{p}$, $i$, and $\phi$, as returned by the fit described in $\S$\ref{SecTransFit}, and the small planet approximation of 
\citet{Mandel}. 

The statistic that has been used to quantify the improvement of the transit
model over the constant flux model is
$\Delta\chi^{2}\%$, the percentage improvement in $\chi^{2}$ of the transit model
($\chi_{T}^{2}$)
over the constant flux model ($\chi_{C}^{2}$). That is
$\Delta\chi^{2}\%$ = $100 \times \frac{ \chi_{C}^{2} - \chi_{T}^{2}}{\chi_{C}^{2}}$.
The anti-transit model is identical
to the transit model except that the transit causes brightening as
opposed to dimming.
The best-fit anti-transit model was determined through the identical method
as outlined in steps $\S$\ref{SecTransBLS} through $\S$\ref{SecTransFit}.
Simultaneously determining the best-fit transit, and anti-transit model results in a minor
increase in computational duration, but this is a justified sacrifice as it provides
an independent method for determining the detection sensitivity.
In a similar fashion, the improvement of the
anti-transit model ($\chi_{AT}^{2}$) over the constant-brightness
model is given by $\Delta\chi^{2}_{-}\%$ = $100* \frac{ \chi_{C}^{2}
- \chi_{AT}^{2}}{\chi_{C}^{2}}$. The quantitative detection criteria
were motivated by \citet{Burke} and have been refined through the
Monte Carlo tests described in $\S$\ref{SecMonte}.

A believable transit was defined as one that satisfied the following criteria:

\begin{itemize}
\item Of the transits selected by the selection criteria ($\S$\ref{SecTransSelect}) the transit of interest showed
the greatest improvement in $\Delta\chi^{2}\%$. 

\item The ratio between the improvement of the transit model over the constant-brightness model versus the anti-transit
over the constant-brightness model is at least \TransAntiBelieve \
($\Delta\chi^{2}\%$/$\Delta\chi^{2}_{-}\%$ $\ge$ $\TransAntiBelieve$).

\item The quantity $f$, as defined by \citet{Burke} and explained below, is less than \Flimit \ ($f < \Flimit$).   
\end{itemize}

The first criterion is intended to be conservative, as it helps
to avoid identifying harmonics of the period of an obvious transit
candidate. The second criterion was suggested by \citet{Burke} for
a groundbased transit survey, and is primarily used as a threshold to 
differentiate astrophysically credible transit
candidates from false positives. We use a much more sensitive value
for the ratio ($\Delta\chi^{2}\%$/$\Delta\chi^{2}_{-}\%$ $\ge$
$\TransAntiBelieve$), due to the increased sensitivity of the {\it MOST}
space-based photometry, as well as the results of the Monte Carlo tests of $\S$\ref{SecMonte}.
The level that has been chosen requires the best-fit transit to be only slightly more prominent than
the best-fit anti-transit candidate.
It is important to note that the best-fit anti-transit is not restricted to at 
the same period as the best-fit transit. The best-fit anti-transit is determined in a completely analogous
fashion to the best-fit transit, and thus the parameters of the best-fit transit and anti-transit are often completely
independent.
The motivation for selecting this sensitive value of \TransAntiBelieve \ was based on
the experience gained from the Monte Carlo results,
as it was the set at that level at which a transit could be reliably recovered from the data-set without inducing
significant numbers of false-positives. The full explanation of how the value of $\Delta\chi^{2}\%$/$\Delta\chi^{2}_{-}\%$ $\ge$
$\TransAntiBelieve$ was derived is given in $\S$\ref{SecMonte}.
The rationale for this cut, as noted by \citet{Burke}, is that in the case that
the light curve shows periodic variability or significant brightening
or dimming trends, a transit search will often assign considerable
significance to these events, possibly resulting in false positive detections. Although this cut has primarily been
designed for wide-field, or open and globular cluster transit searches, where
eliminating targets that show obvious sinusoidal variation is
desirable, it is still applicable to the present search to account
for possible variations and trends due to stray earthshine or
intrinsic variability in the star HD 209458. It should also be noted that only
improvements in the anti-transit model, $\Delta\chi^{2}_{-}\%$ ($\Delta\chi^{2}_{-}\%$ $>$ 0),
are accepted. This is because, in the limit of extremely deep transits, on the order of those visible by eye
such as those produced by HD 209458b, the 
anti-transit statistics are forced to return worse values of $\Delta\chi^{2}_{-}\%$, as no significant
brightening event is detected.
This is because the median rather than the mean is used for
the out-of-transit component of our transit model.
This effect was not observed in the current application, or 
in the Monte Carlo tests, but it is a useful caveat for other transit searches.

The third criterion invokes the parameter $f$, which \citet{Burke}
defines as $f$=$\chi^{2}_{kth}/\chi^{2}_{total}$. Here $\chi^{2}_{k}$
refers not to the typical $\chi^{2}$, but rather to summing the
following quantity for the $k^{th}$ transit: $\chi^{2}_{kth}$ =
$(m_{i}^{2}/\sigma_{i}^{2})$, where $m_{i}$ and $\sigma_{i}$ are the
magnitude and uncertainty, respectively, of the $i^{th}$ measurement
during the $k^{th}$ transit. The criterion requiring $f < \Flimit$
roughly corresponds to observing the transit at least one and a half
times, assuming similar noise. This criterion is useful in eliminating
long-period false-positive events, as it ensures that the dimming
behaviour that caused the EEBLS routine to flag the event as highly
significant, is actually observed at least one and a half times. 
This cut is of only limited significance
to our HD 209458 search, given the sensitivity, time span, and high
duty cycle of the {\it MOST} photometry. However it remains a safeguard for some long-period events in our current search
and for future ground and space-based searches.

Avoiding false positives is necessary 
so as to not place unjustifiably sensitive limits in the Monte Carlo statistics of $\S$\ref{SecMonte}.
These or similar criteria should be suitable for other ground and space-based transiting exoplanet searches.

% It is our opinion that these detection criteria are sufficient for {\it MOST} photometry.
\section{Monte Carlo statistics}
\label{SecMonte}

To assess the sensitivity of the aforementioned search routine and the {\it MOST} data set
to other transiting planets in the HD 209458 system, simulated transits for planets of various radii
and orbital parameters were inserted 
into the {\it MOST} photometry and Monte Carlo statistics of the
transit recovery rate were generated.

\begin{figure}
\epsscale{0.6}
\includegraphics[angle=270, scale = 0.52]{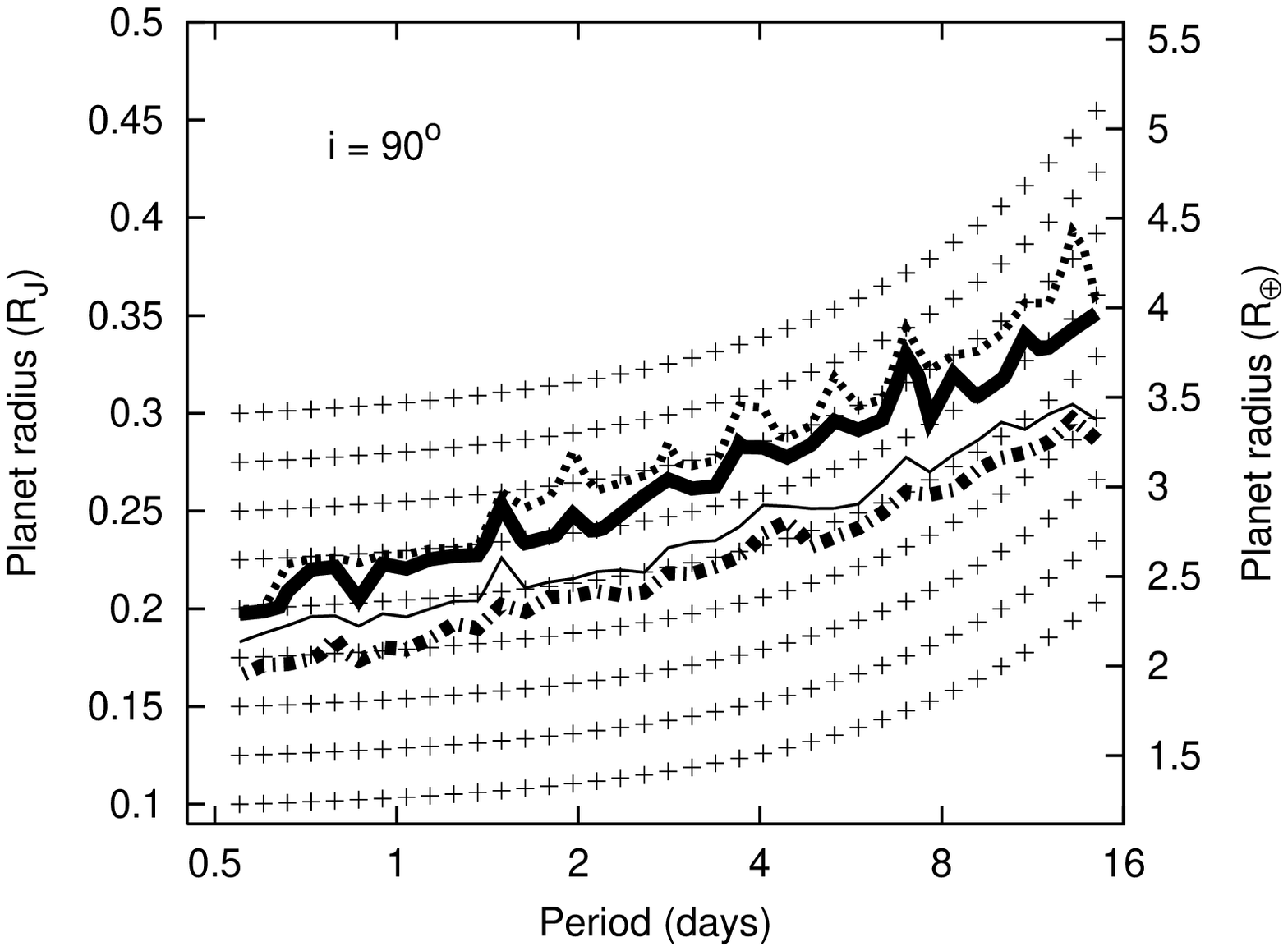}
\includegraphics[angle=270, scale = 0.52]{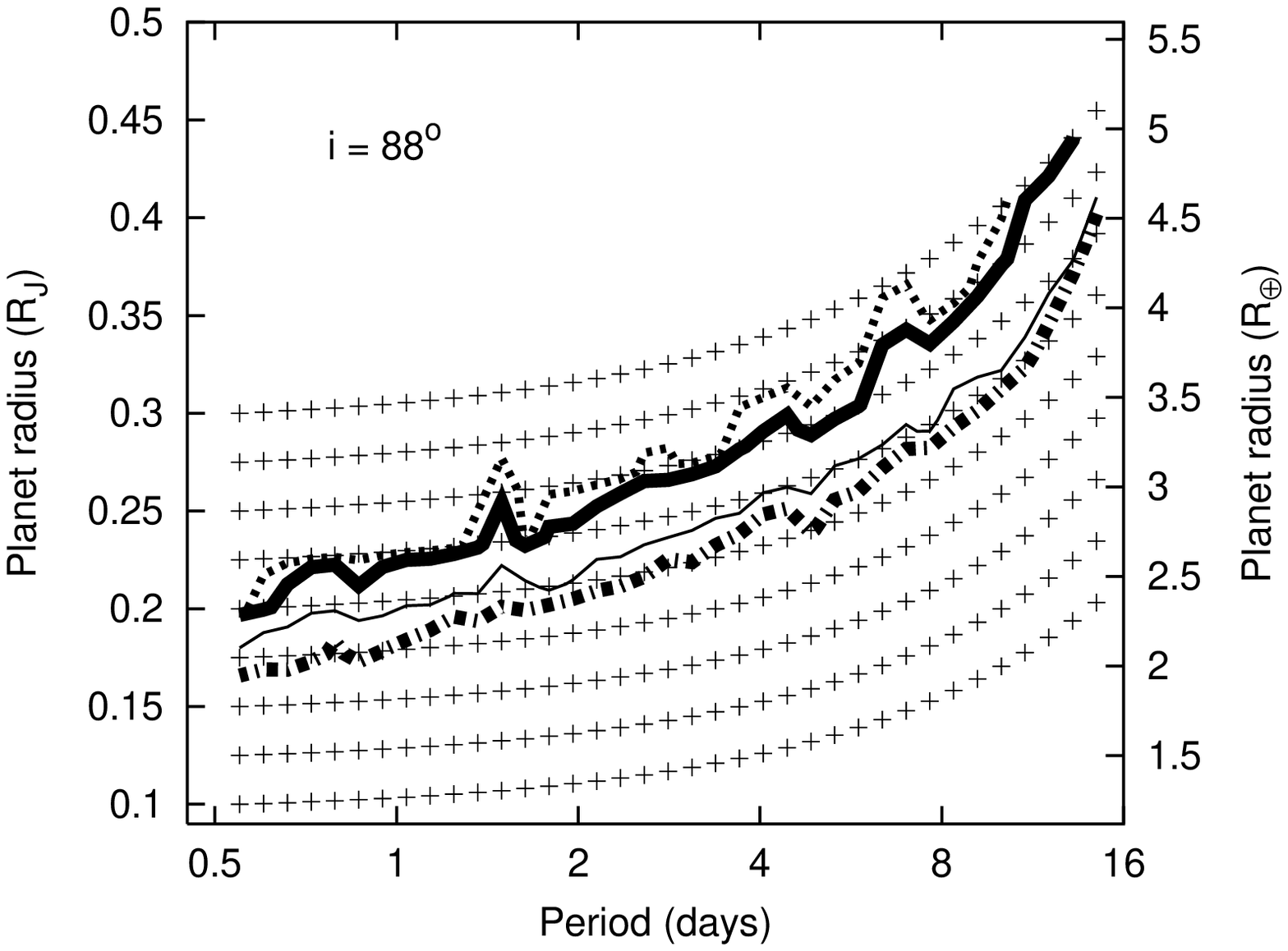}

\includegraphics[angle=270, scale = 0.52]{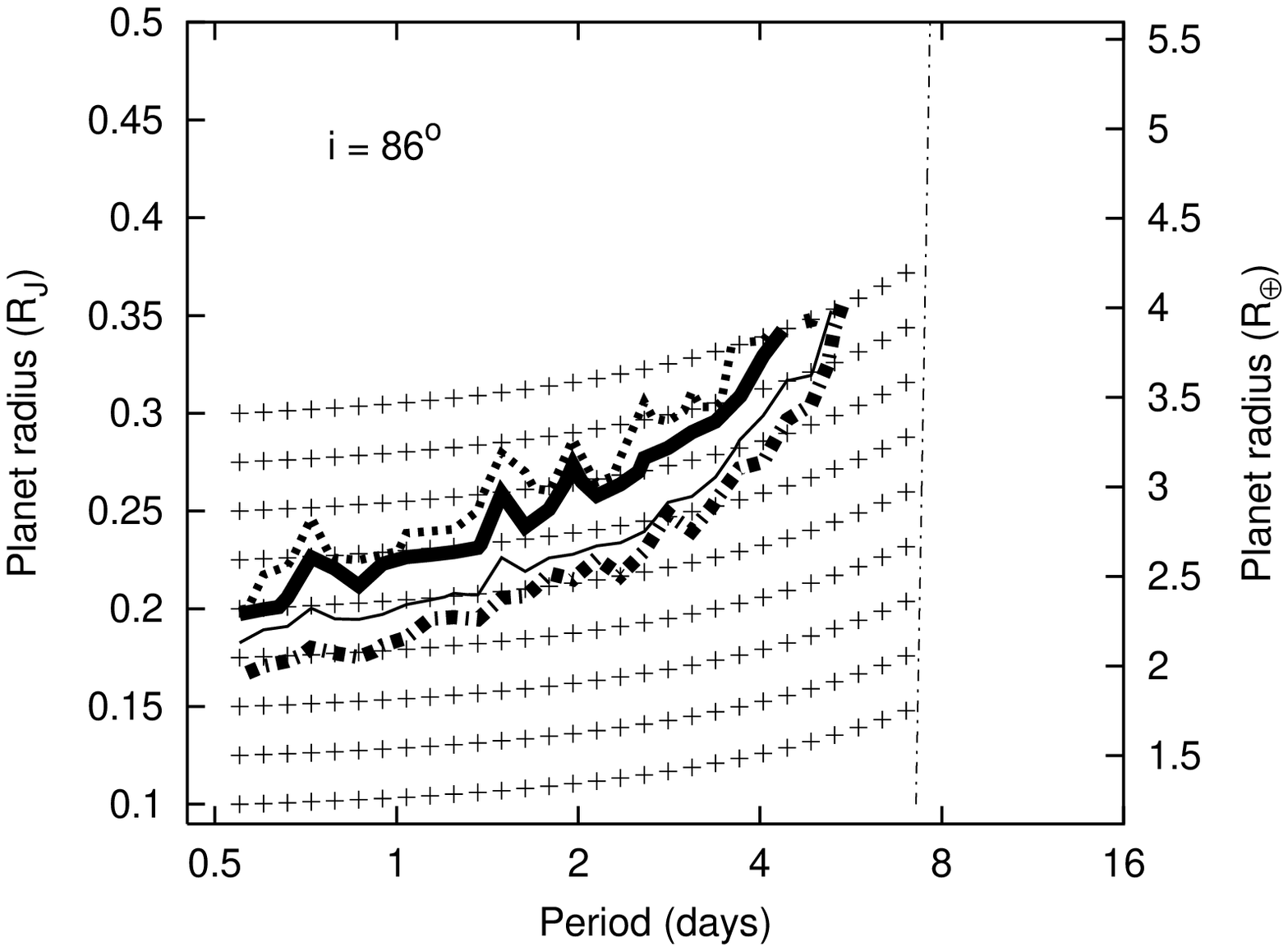}
\includegraphics[angle=270, scale = 0.52]{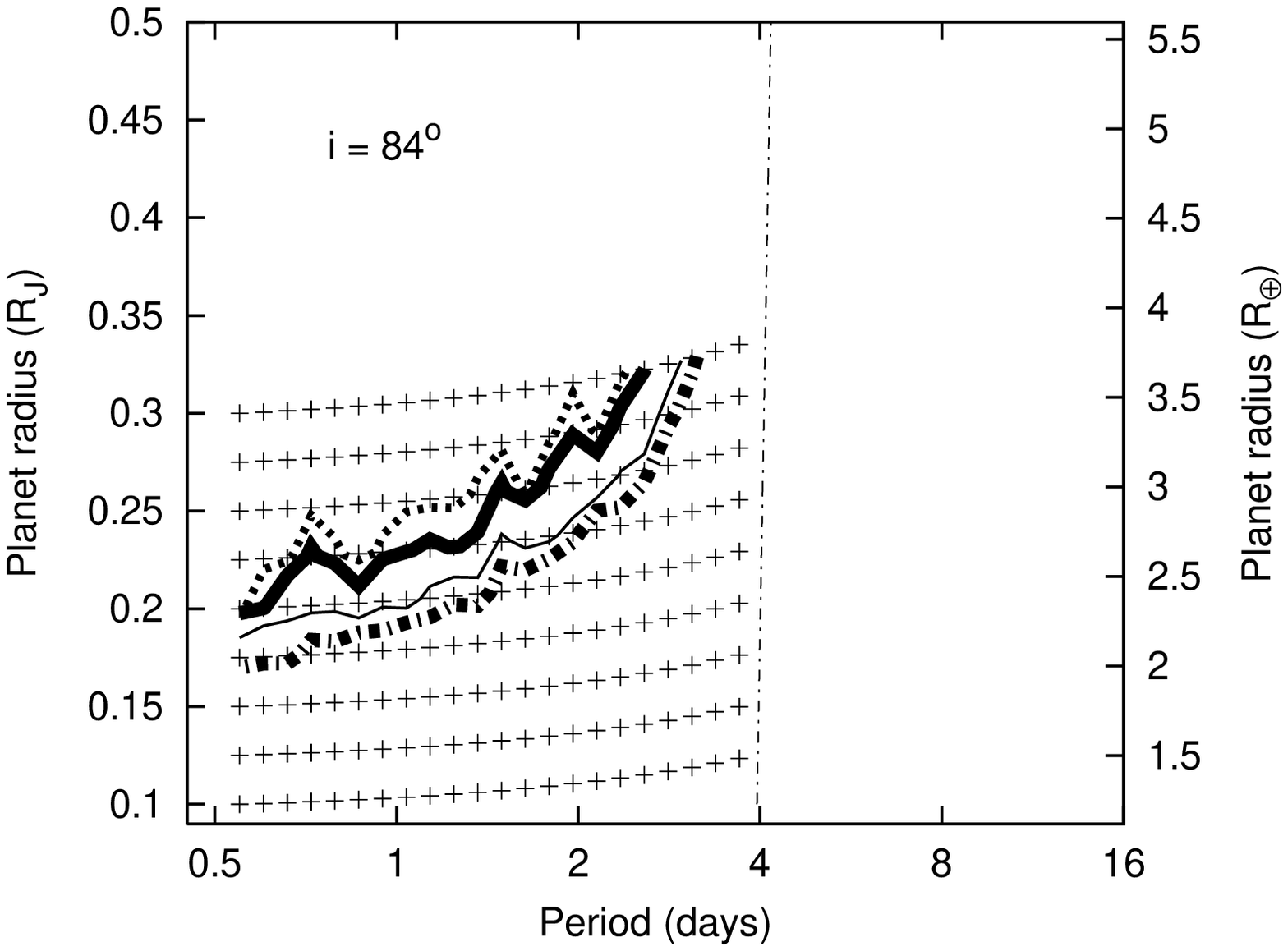}

\includegraphics[angle=270, scale = 0.52]{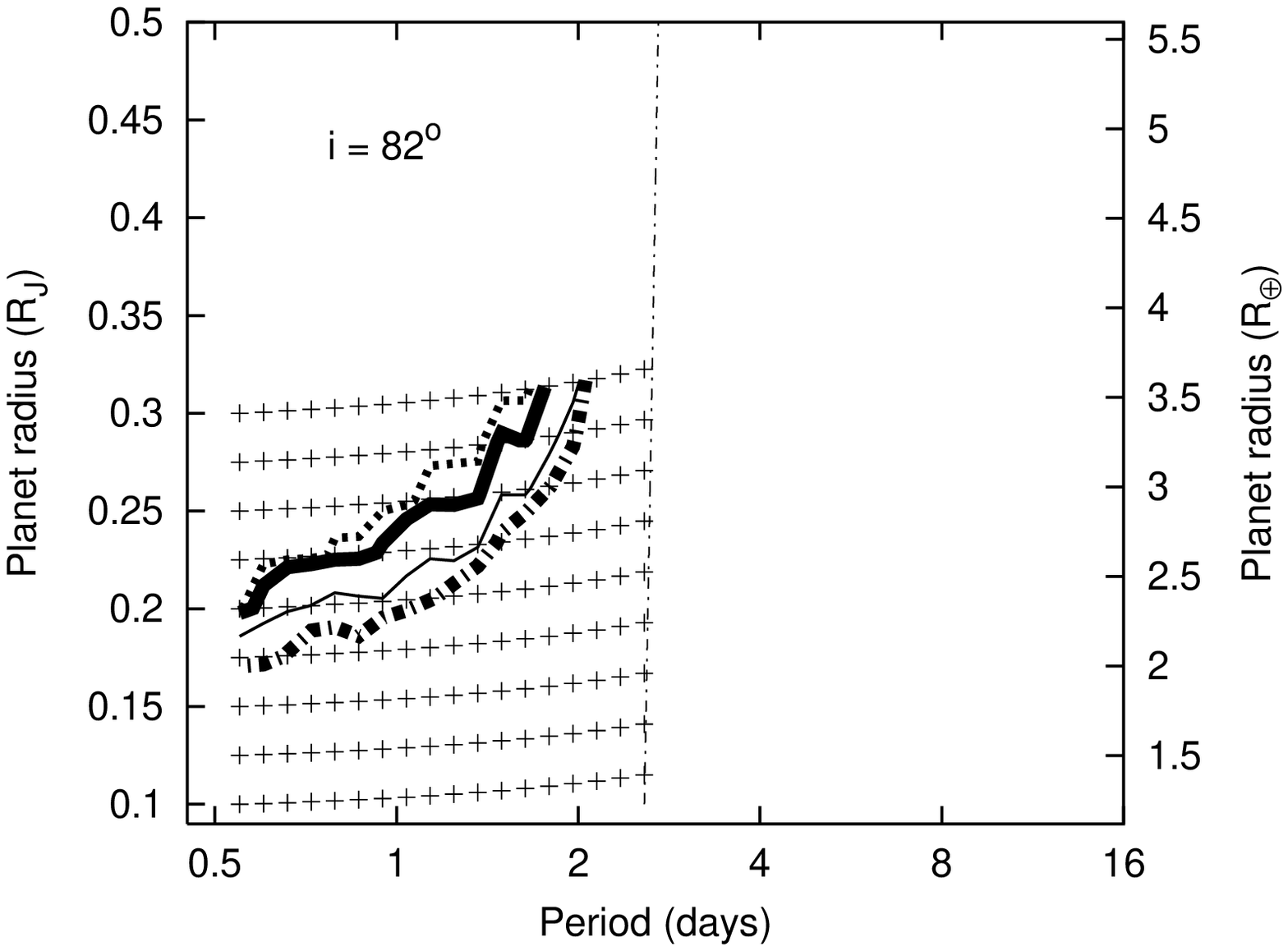}
\includegraphics[angle=270, scale = 0.52]{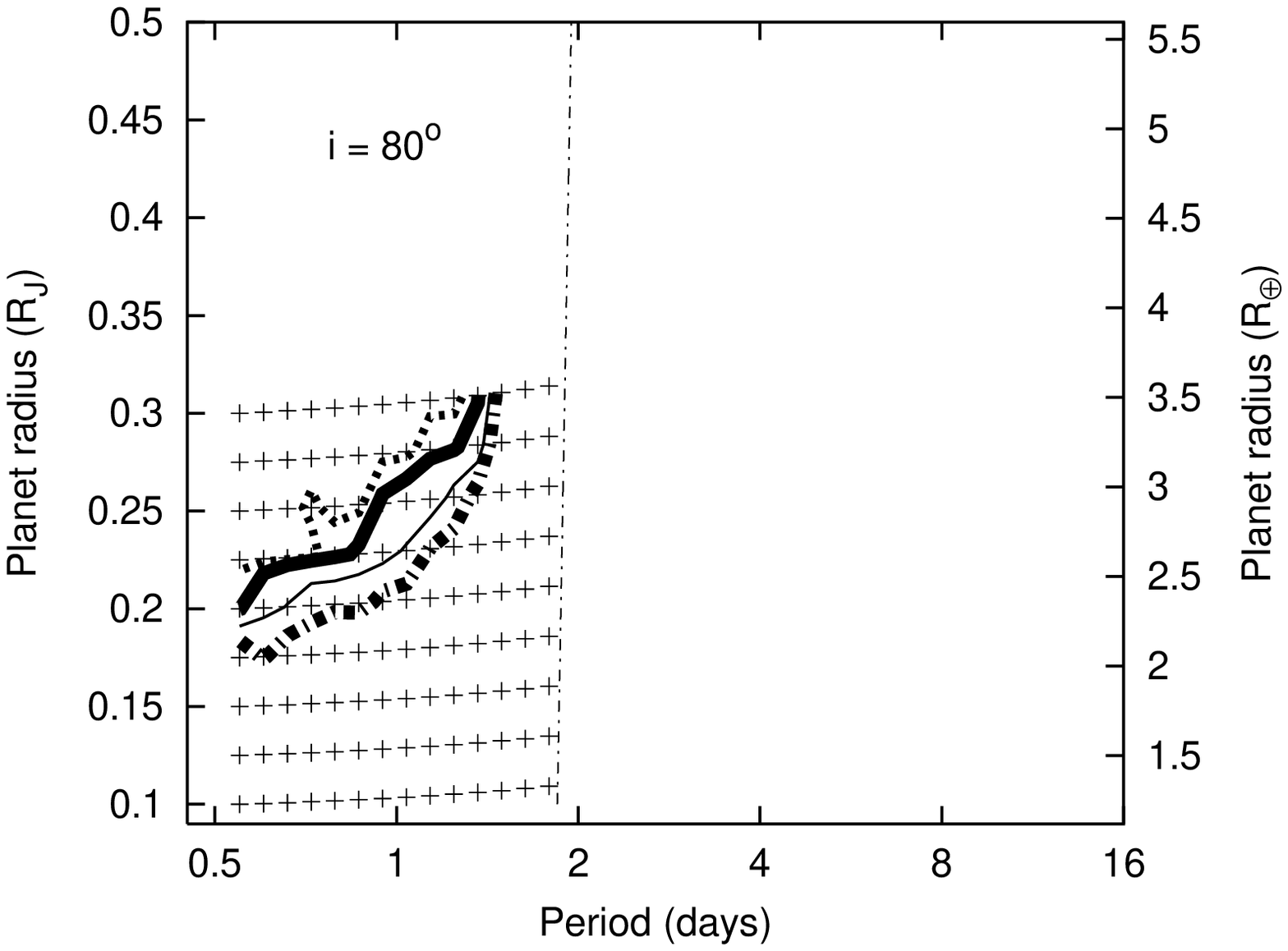}

\caption{
	{Confidence limits in transit detection as a function of planet radius and planetary orbital period, for different orbital
	inclinations, based on Monte Carlo statistics. The crosses represent the radii and periods at which synthetic
	transits were inserted in the data.
	The dotted, thick-solid, and thin-solid lines represent the 99, 95 and 68\%
	confidence
	contours, respectively.
	The thick dot-dashed curve represents the 68\% confidence contour if the criterion $\Delta\chi^{2}\%$/$\Delta\chi^{2}_{-}\%$ $\ge$ \TransAntiBelieve \
	was not used.
	The near vertical dot-dash line in the later panels indicates 
	the maximum period that produces a transit for that given inclination angle.
	Note the logarithmic
	period scaling on the x-axis. 110 phases were inserted for the $90^{o}$
	inclination angle case, while 65 phases were inserted for the other inclination angles. 
	}  
	\label{FigMonte}
}
\end{figure}

\begin{figure}
\epsscale{0.6}
\includegraphics[angle=270, scale = 0.55]{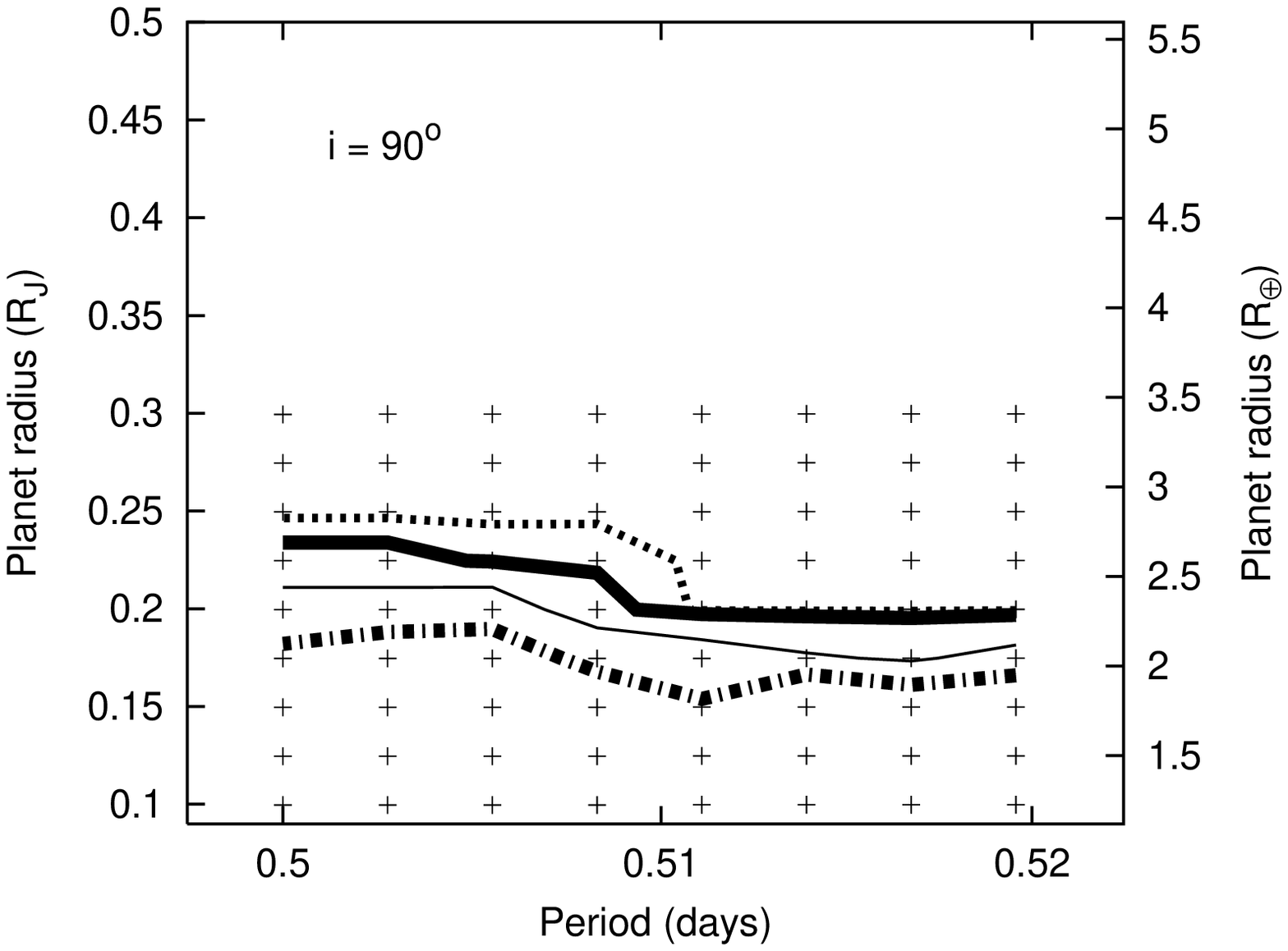}
\includegraphics[angle=270, scale = 0.55]{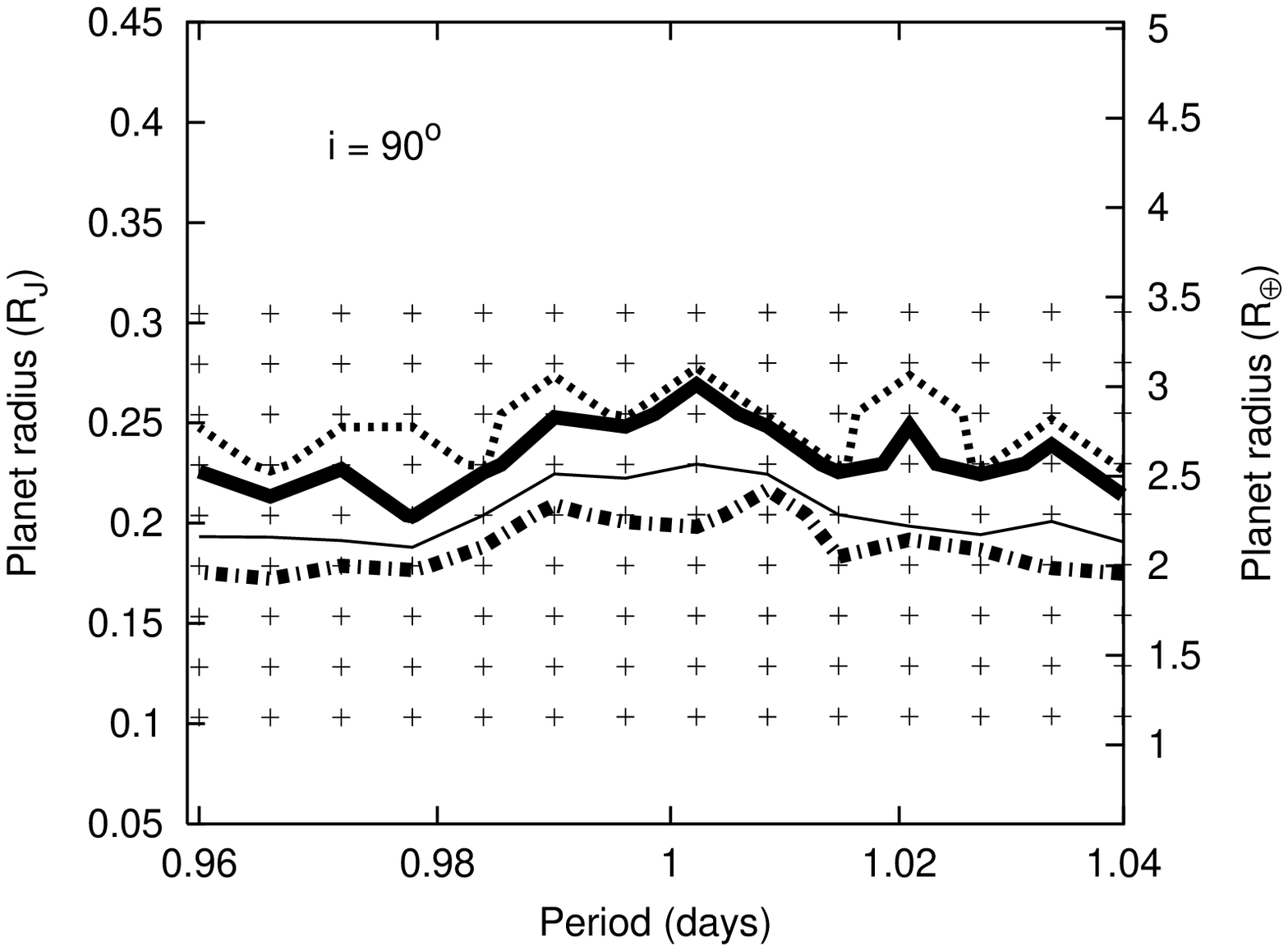}

\caption{
	{Transit search sensitivity near the periods of 0.5 (left) and 1.0$d$ (right), 
	where sinusoidal terms were filtered
	from the photometry. 
	The format is the same as Fig. \ref{FigMonte}. There were 25 phases inserted for each period-radius point.
	As can be seen the transit search sensitivity is only slightly adversely affected near these periods. 
	}  
	\label{FigCutOut}
}
\end{figure}

Realistic limb-darkened transits due to planets with various radii
$R_{p}$, orbital phases $\phi$, periods $P$, and inclinations $i$,
were inserted into the 2004 and 2005 {\it MOST} HD 209458 data. The small planet approximation of
\citet{Mandel} was used. The differences between the more
accurate non-linear model of \citet{Mandel} and the small planet model
approximation are negligible for the size of planets inserted in this data-set, even for transits with short periods,
and thus semi-major
axes only a few times the stellar radius. 
These
modified data were then subjected to the analysis described above.
Transits were inserted with logarithmic period spacing (as discussed
in $\S$\ref{SecTransBLS}), with $\eta_{inp} = 0.095$ in the period
range $0.55 d < P_{inp} < 14.5 d$. In total, 37 period steps were
used for the 90$^{o}$ and 88$^{o}$ inclination angle cases. For other inclination angles transits were inserted with logarithmic
period spacing until the period exceeded 
the maximum period, or semi-major axis, that would produce a transit, a $<$ $(R_{*}+R_{P})$/cos$i$. 
For each trial period, simulated transits corresponding to 9
different exoplanet radii were inserted, sampling the period-radius
space of interest. For each period and radius, 110, 65, or 25 phases were
inserted as summarized in Figures \ref{FigMonte} and \ref{FigCutOut}.
For each of these points, the phase $\phi$ was generated
randomly to be in the range 0 $\le$ $\phi$ $<$ 1.
Because a 2.4 Ghz Pentium processor with 1 Gbyte of memory can perform
the transit search algorithm on an individual MOST HD 209458 data set
in $\sim$10 minutes, exploration of the entire grid just mentioned for
all inclinations involves $\simeq10^5$ iterations and 2.0 CPU years.
The calculation was performed on the LeVerrier Beowulf cluster in
the Department of Physics and Astronomy at the University of British
Columbia using 45 dual-CPU compute nodes.

An inserted transit
was judged to be detected if the parameters $\phi$ and $P$ returned
by the transit search algorithm were sufficiently close to the input
values, $\phi_{inp}$, $P_{inp}$. The returned period had to satisfy
the following criteria: $|P - P_{inp}| < 0.1 d$ and $|P/P_{inp} - 1| <
1\%$. The limits of the criterion for $\phi$ were dependent on the orbital period,
because as the period of the putative transiting planet decreases,
the fractional transit length increases
accordingly. Thus the accuracy required in the determination of
$\phi$ was relaxed for shorter periods. The criterion on $\phi$
is: $|\phi-\phi_{inp}|$ $<$ $0.09 - 0.0054$($P_{inp}$ - 0.5$d$)/$d$. Obvious multiples of the period of the inserted 
planet, up to $4$ times the inserted period, $P_{inp}$,
as well as half-period ($P \approx \frac{1}{2}P_{inp}$), and one-third period ($P \approx \frac{1}{3}P_{inp}$)
solutions as returned by the transit search algorithm ($\S$\ref{SecTrans}) were also
accepted, as it was found for a low percentage of cases that harmonics or subharmonics of the inserted trial period were flagged
as the best candidates. For the 90-degree inclination angle the best candidates had a period double, triple, or quadruple
the inserted period approximately 1.0\%, 0.5\%, and 0.3\%, respectively, 
of the total transit detections in the Monte Carlo analysis, while half and one-third
period solutions accounted for 0.3\% and 0.1\% of the aforementioned total, respectively.
 
These Monte Carlo results were also used to determine the level above which a transit recovered from the data-set could
be considered significant. In deriving this value one must be careful to closely balance the desire to use as low, and thus
sensitive, a
limit as possible with the necessity of limiting the number of false positive detections.
Thus, to satisfy this rationale the following method
has been used to determine this value of $\Delta\chi^{2}\%$/$\Delta\chi^{2}_{-}\%$ that will denote a believable transit.
To determine a suitable value of the transit
over the anti-transit model ratio, $\Delta\chi^{2}\%$/$\Delta\chi^{2}_{-}\%$,
all Monte Carlo cases where the candidate returned by the transit routine was not the candidate that was inserted 
were investigated.
These events where the candidate returned by the routine was not the inserted transit will 
henceforth be referred to as spurious events.

By initially investigating the Monte Carlo statistics without the criterion that 
$\Delta\chi^{2}\%$/$\Delta\chi^{2}_{-}\%$ $\ge$ $\TransAntiBelieve$
we were able to investigate the maximum efficiency of the current transit routine. By comparing the efficiency of the Monte Carlo
results with and without this threshold one can quantitatively investigate the impact of this criterion. 
We restrict the investigation of these spurious signals to those that the routine has a reasonable chance of actually detecting.
This area was determined to be the level above approximately the 25\% contour of the Monte Carlo statistics when this threshold criterion was not
used. Thus for the generation of spurious signal statistics we only use 
those Monte Carlo cases with an input radius above the limit given
here: $R_{Pinp} > [0.125 + (0.1/14.7)(P_{in}-0.5days) )$/$days$ $R_{J}$.

\begin{figure}
\includegraphics[height = 4.8in, width = 3.2in, angle=270]{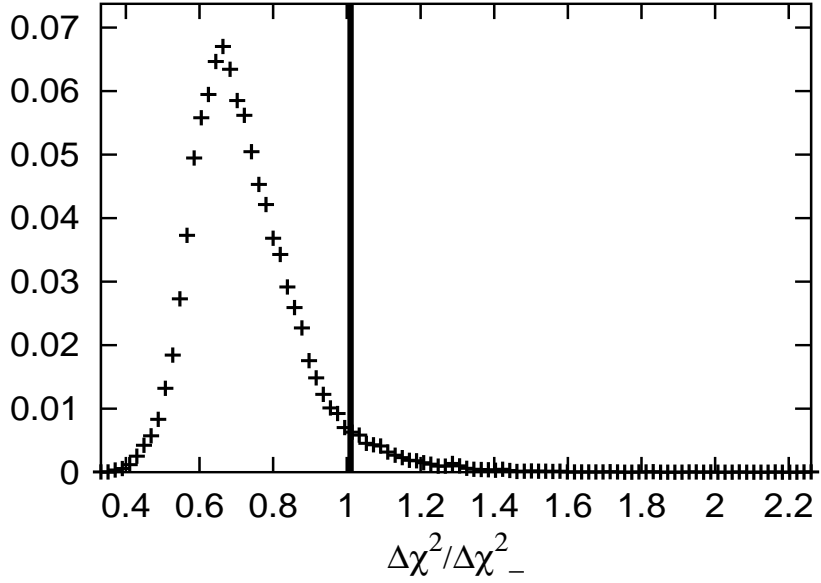}
\caption{	
	{A histogram showing the improvement of the transit over the anti-transit model ($\Delta\chi^{2}\%$/$\Delta\chi^{2}_{-}\%$)
	for all Monte Carlo candidates that are spurious (when the transit recovered by the transit routine is not the inserted transit).
	The threshold level for $\Delta\chi^{2}\%$/$\Delta\chi^{2}_{-}\%$ above which
	a transit can be considered significant is determined via these statistics as it is set 
	at a level, $\Delta\chi^{2}\%$/$\Delta\chi^{2}_{-}\%$ $\approx$ \TransAntiBelieve, 
	that would rule out 95\% of these spurious signals. The solid-vertical line thus 
	marks this threshold value.
	}  
	\label{FigHisto}
}
\end{figure}

A histogram
of these spurious events versus $\Delta\chi^{2}\%$/$\Delta\chi^{2}_{-}\%$ for all inserted candidates above 
the radius threshold discussed can
be seen in Figure \ref{FigHisto}.
It was decided that a suitable level of $\Delta\chi^{2}\%$/$\Delta\chi^{2}_{-}\%$ would be one that ruled out 95\% of these 
spurious events.
Thus the threshold level of $\Delta\chi^{2}\%$/$\Delta\chi^{2}_{-}\%$ $\ge$ \TransAntiBelieve \ has a transit over anti-transit ratio
greater than 95\% of these spurious signals.
That is, if there is a transiting planet observable 
in the HD 209458 system, other than the known exoplanet HD 209458b, 
and the routine returns a putative transit that exceeds
the threshold $\Delta\chi^{2}\%$/$\Delta\chi^{2}_{-}\%$ $\approx$ \TransAntiBelieve,
one can be at least 95\% certain that an actual transiting planet has been recovered, rather
than a spurious signal. 
Since this measure has been
calibrated to this MOST data set, the length of the time series has
already been accounted for.

The fractions of times that the artificially inserted transits were
recovered from the data for various radii, periods, and inclination
angles are given in Figure \ref{FigMonte}. Also indicated in Figure \ref{FigMonte} is the 68\% contour limit
that would be placed without the $\Delta\chi^{2}\%$/$\Delta\chi^{2}_{-}\%$ $\ge$ \TransAntiBelieve \ 
criterion.
The close agreement between the 68\% contour with or without this criterion ($\Delta\chi^{2}\%$/$\Delta\chi^{2}_{-}\%$ $\approx$ \TransAntiBelieve)
indicate that this criterion does not significantly impact the sensitivity of the Monte Carlo statistics generated, while providing
a robust limit so as to avoid false positives and spurious detections.

\begin{figure}
\epsscale{0.6}
\includegraphics[angle=270, scale = 0.54]{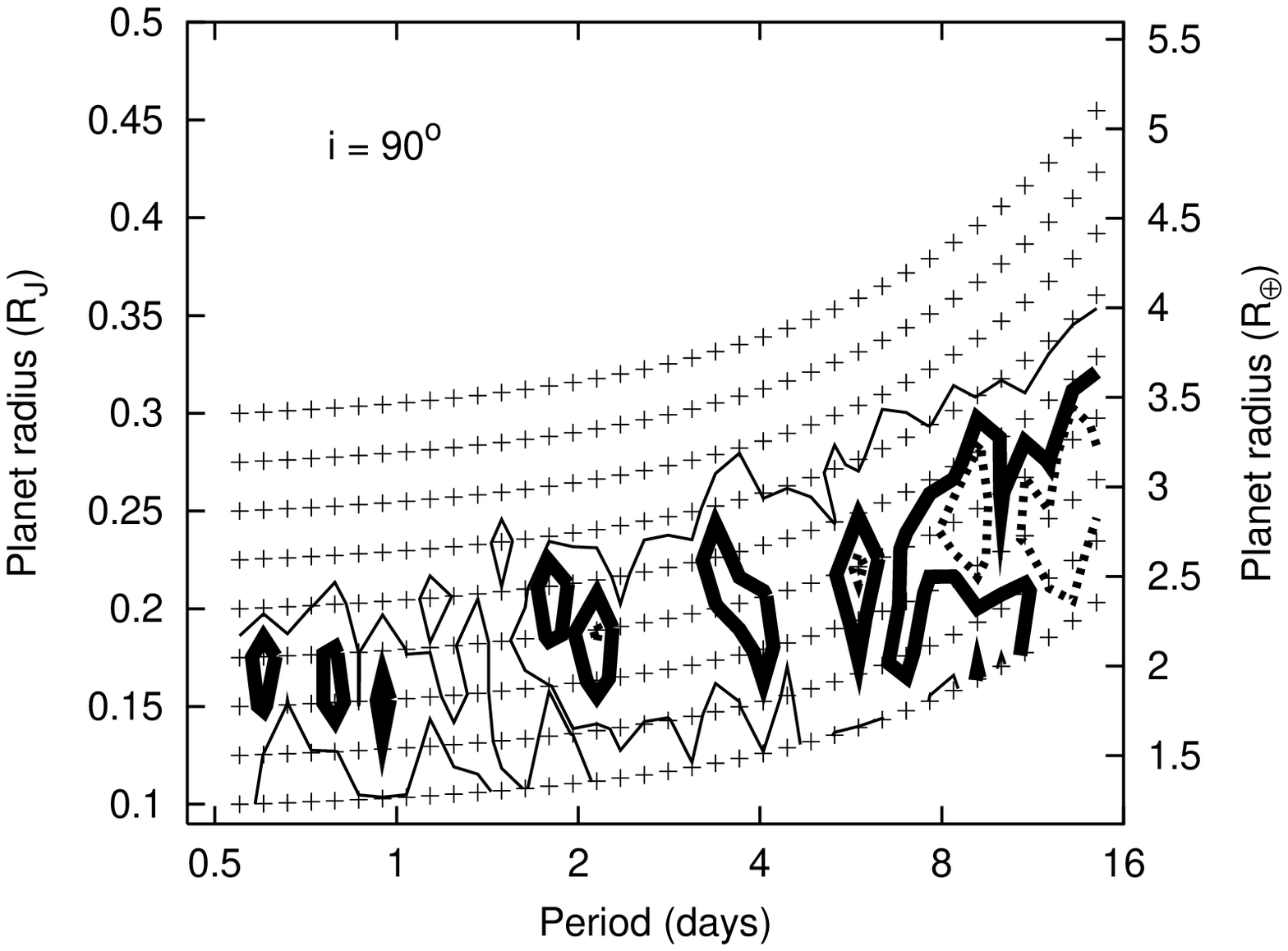}
\includegraphics[angle=270, scale = 0.54]{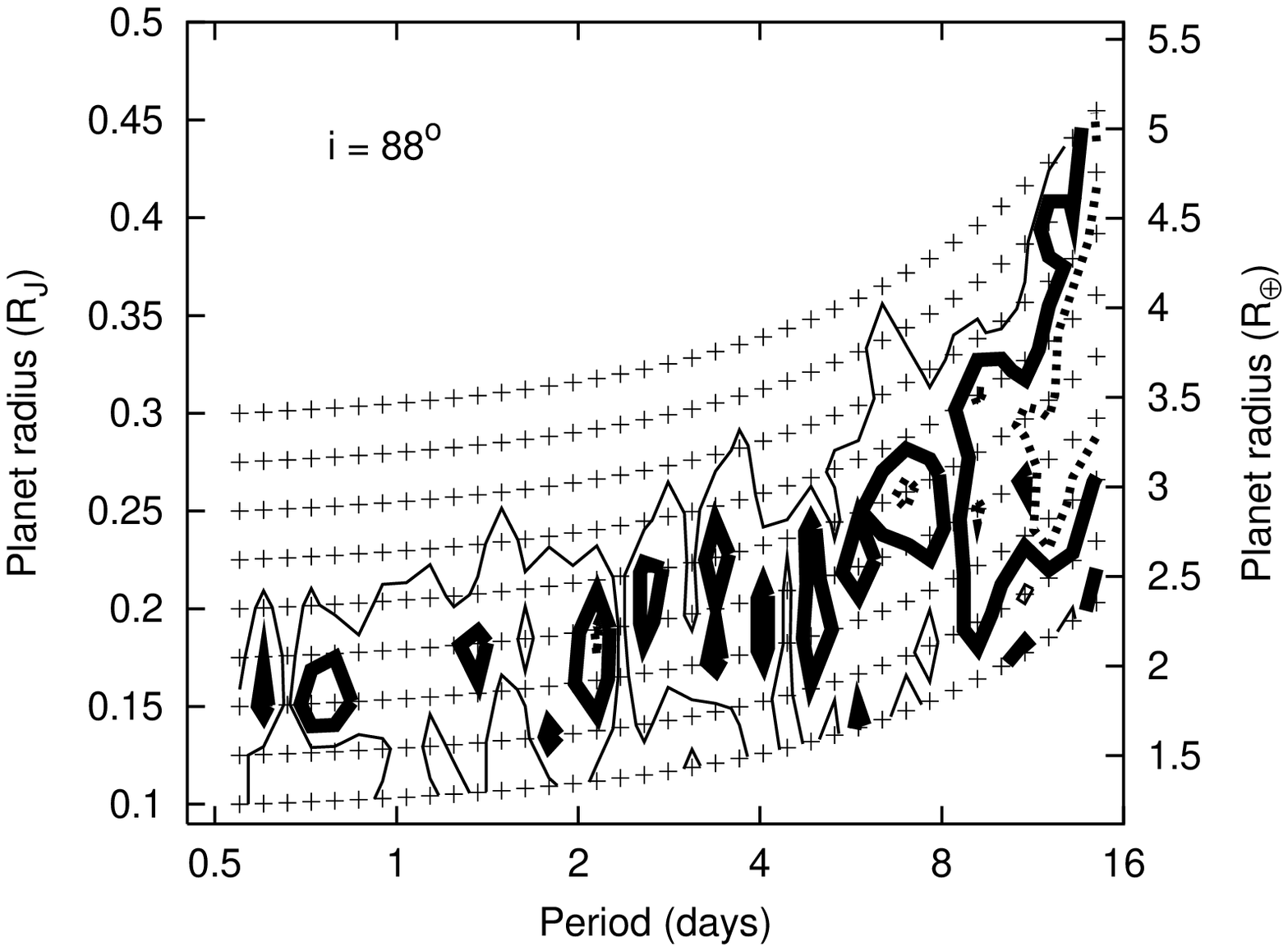}

\includegraphics[angle=270, scale = 0.54]{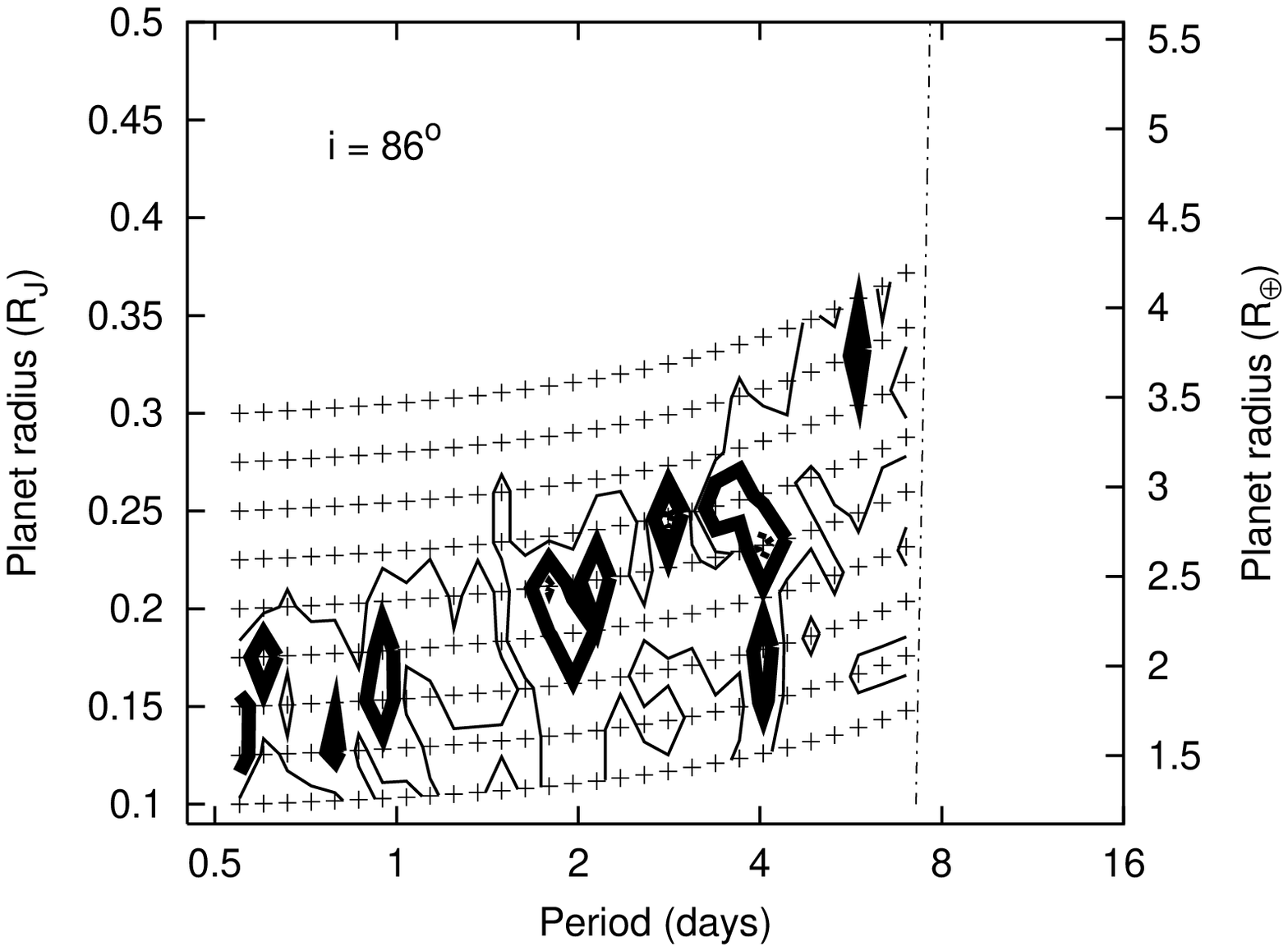}
\includegraphics[angle=270, scale = 0.54]{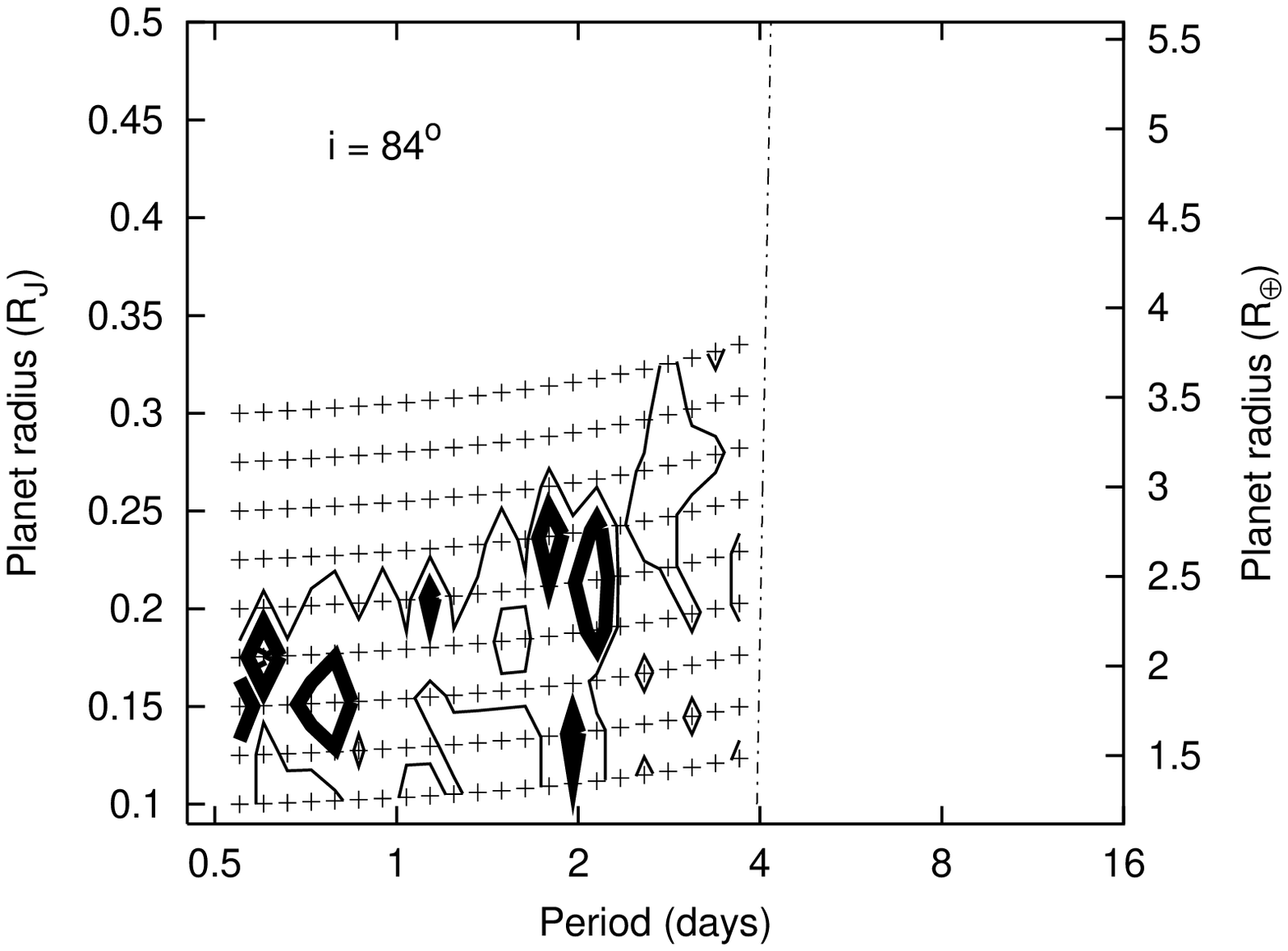}

\includegraphics[angle=270, scale = 0.54]{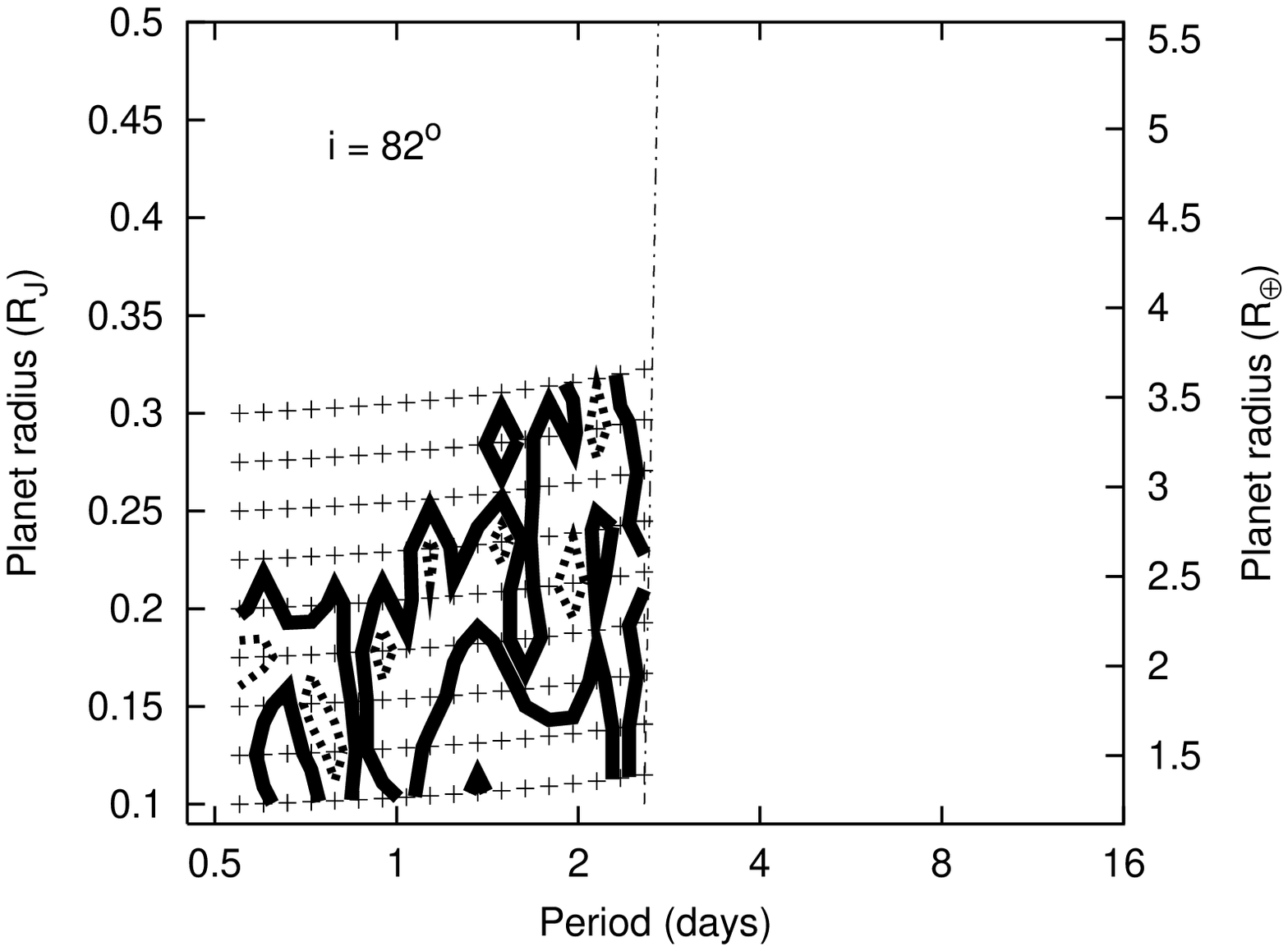}
\includegraphics[angle=270, scale = 0.54]{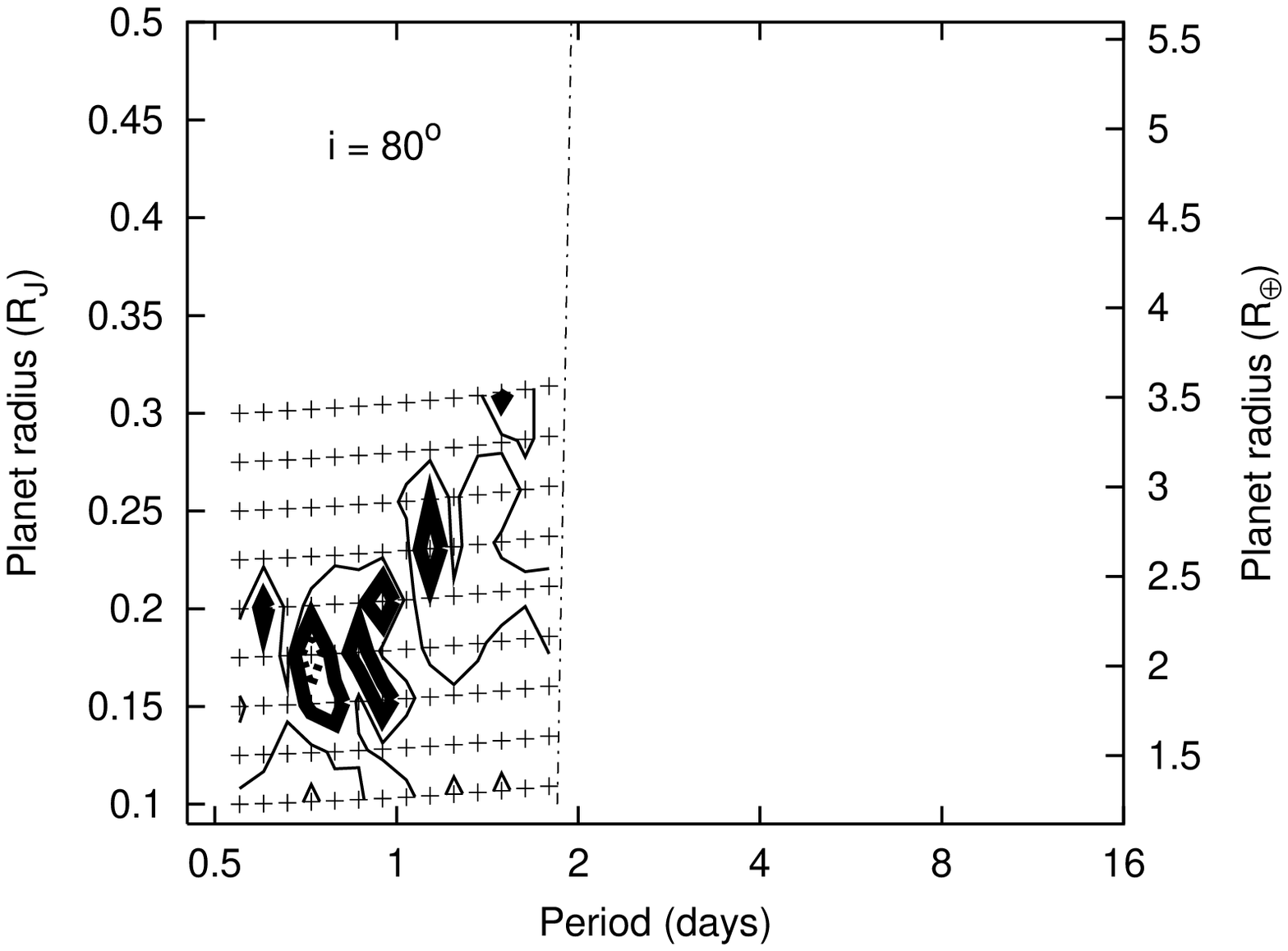}

\caption{
	{Likelihood of spurious transit detections returned by the Monte Carlo statistical analysis. The
	dotted, thick-solid, and thin-solid lines represent the 10, 5 and 1\% spurious signal contours, respectively.
	The format of the figure is otherwise identical to Figure \ref{FigMonte}. Note that the 
	spurious signals occur in the intermediate regions of the period-radii space of interest,
	where it is not guaranteed that the correct transit will be recovered,
	but the inserted transit still causes significant deviations to the light-curve.
	}  
	\label{FigMonteFalse}
}
\end{figure}

We also explore the distribution of spurious signals across period-radii space.
These values are displayed in Figure \ref{FigMonteFalse}, and indicate that spurious signals
are negligible (below 1\%) other than the intermediate areas where the routine
has a small to moderate chance to correctly determine
the inserted transit.
In these intermediate areas in period-radius space
the inserted transit is small enough that correctly identifying the period, radius,
and phase of the inserted transit using this routine is not guaranteed. However, the inserted transit
is of sufficient magnitude to significantly impact the shape of the light-intensity
curve and lead to spurious transit signals rising above the threshold, $\Delta\chi^{2}\%$/$\Delta\chi^{2}_{-}\%$ $\ge$ \TransAntiBelieve,
for a small, but notable, percentage of cases. Accurately determining the frequency of true false positives - where there is not another transiting
planet in the data-set - as opposed to spurious signals is not attempted in this application,
as it is difficult to produce a synthetic data-set that
accurately reproduces the systematics of the real data-set.
These systematics may be the result of imperfect removal of stray earthshine from the {\it MOST} data,
or may, in fact, be intrinsic to the star. 

We also investigated the performance of our routine at harmonics and subharmonics of the periods of the sinusoids that 
were removed from the data ($P$=1.0$d$, and $P$=0.5$d$). Due to the fact that 
the fractional length of a transit is comparatively large for short periods of 0.5 - 1.0 $d$ it was found 
to be difficult to
properly recover a transit signal near the harmonics and subharmonics of these periods for all cases.
As expected and shown in Figure \ref{FigCutOut}, the routine performs
minimally worse near these values as a portion of the
transit signal is removed when the sinusoidal fit is removed
in $\S$\ref{SecTransReduce}. As can be seen in Figure \ref{FigCutOut},
however, the period range demonstrating these drops in survey sensitivity
is very narrow ($\frac{ \Delta P }{ P }$ $\approx$ 2\%), and the effect is very minor. Therefore very sensitive
limits can still be set at these periods.

The harmonics
and subharmonics of the period of the known planet HD 209458b, $P$ $\approx$ $3.5247$ \citep{Knutson},
are also expected to show decreased sensitivity to transits 
due to the fact that the transits of the known planet were excised from the data. 
This reduction of sensitivity
at these periods is unfortunate considering that it is expected that low-mass planets show a preference for 
these resonant orbits (\citealt{Thommes}; \citealt{Zhou}). 
The fractional length in phase that was removed at $P$ $\approx$ $3.5247$ for the transit was 4.1\%
of the total.
Thus for the subharmonics of the known planet ($P$ $\approx$ 7.05, 10.6, 14.1$d$) in 
approximately 4.1\% of the
cases the transits will coincide with the transits of the known planet, and thus the 
routine will be unable to recover the transiting planet, because
this data will have been completely removed.
For these 
periods the 99\% and 95\% contours of Figure \ref{FigMonte} would likely be seriously affected,
although the 68\% contour
should only be slightly affected.
For
the harmonics of the orbital period of HD 209458b, the
situation is better, since not all of the transit events
would be removed from the data.
For periods near the first harmonic ($P$ $\approx$ 1.76$d$), every second occurrence of a transit will be lost
in only 8.2\% of the cases. For the second harmonic ($P$ $\approx$ 1.17$d$),
every third transit would be missed in 12.3\% of the cases.
For periods near these values the sensitivity limits can be expected to be 
only slightly degraded from those shown in Figure \ref{FigMonte}. Even for exact harmonics 
the routine's sensitivity should be only marginally worse than the limits quoted in Figure \ref{FigMonte},
as the remaining transits should allow the correct period (or a multiple of the period) to be recovered.
These resonances have already been sensitively examined using the transit-timing technique
on the known exoplanet HD 209458b (\citealt{Agol}; \citealt{Miller}). 
\citet{Miller} have applied this technique to the {\it MOST} photometry of HD 209458 and are able to rule out sub-Earth-mass planets 
near the inner harmonics of HD 209458b, and Earth and super-Earth mass planets near the outer subharmonics out to the 4:1 resonance.

The detection limits presented here are valid for circular orbits, but are largely applicable to orbits
of other eccentricities.
Eccentric orbits will increase, decrease, or have little effect on the fractional length of the transit, 
depending on whether the transit occurs near apastron, periastron, or in between, respectively.
Obviously, the cases where the fractional transit length is increased will favourably affect the detection limits presented,
while the opposite cases will adversely affect the detection limits.
For low-eccentricity orbits this effect should be negligible, and thus the detection limits presented here should be largely
applicable to these orbits. 
For non edge-on inclination angles, highly eccentric orbits will adversely affect the limits presented,
as a greater fraction of the time the planet will not transit the star along our line of sight.
For edge-on, high-eccentricity orbits the limits will likely be modestly, adversely affected due to the cases where the transit occurs near
periastron. 
Numerical simulations indicate that we would expect 
hot Earths or other exoplanets to have negligible eccentricities in the period range we are investigating \citep{Raymond}.

Thus the search routine described above in $\S$\ref{SecTrans}
should be able to detect planets with radii greater than the limits
given in Figure \ref{FigMonte}. In the most optimistic case of an edge-on $90^{o}$ inclination angle transit, for planets
with periods of approximately a half-a-day to two weeks this limit is
approximately \Rjupsmall \ to \Rjuplarge \ $R_{J}$, or \Rearthsmall \ to \Rearthlarge \ $R_{\earth}$, respectively,
with 95\% confidence.
If one assumes a mean density of $\rho$ $\approx$ 3000 kg m$^{-3}$ 
- a value averaging the various extrasolar super-Earth models discussed below -  
their respective masses would be \Massjupitersmall \ and \Massjupiterlarge \ $M_{J}$,
or \Massearthsmall \ and \Massearthlarge \ $M_{\earth}$, respectively. 
If these putative planet were gaseous, and thus possible hot Neptune analogues ($\rho$ $\approx$ 1600 kg m$^{-3}$),
a planet of $R_{P}$ = \Rearthlarge \ $R_{\earth}$ would result in a planetary mass of \MassearthlargeNeptune \ $M_{\earth}$.
Thus the mass-period parameter space
we have ruled out in this study, assuming these hypothetical mean densities,
sets even tighter limits than those set by radial velocity observations of the system.
\citet{Laughlin} 
for instance ruled out other planets in the HD 209458 system with $M$sin$i$ $>$ 0.3$M_{J}$ for $P$ $<$ 100$d$.
The planets we are thus able to rule out are one-fifteenth to one-third the mass of these radial
velocity limits for edge-on transits. Full discussion of the mass of planets that have been ruled out in this study
is given in $\S$\ref{SecMass}.

\section{HD 209458 transit search}
\label{SecHDTrans}

\begin{figure}
\includegraphics[width=2.3in, height = 6.0in, angle=270]{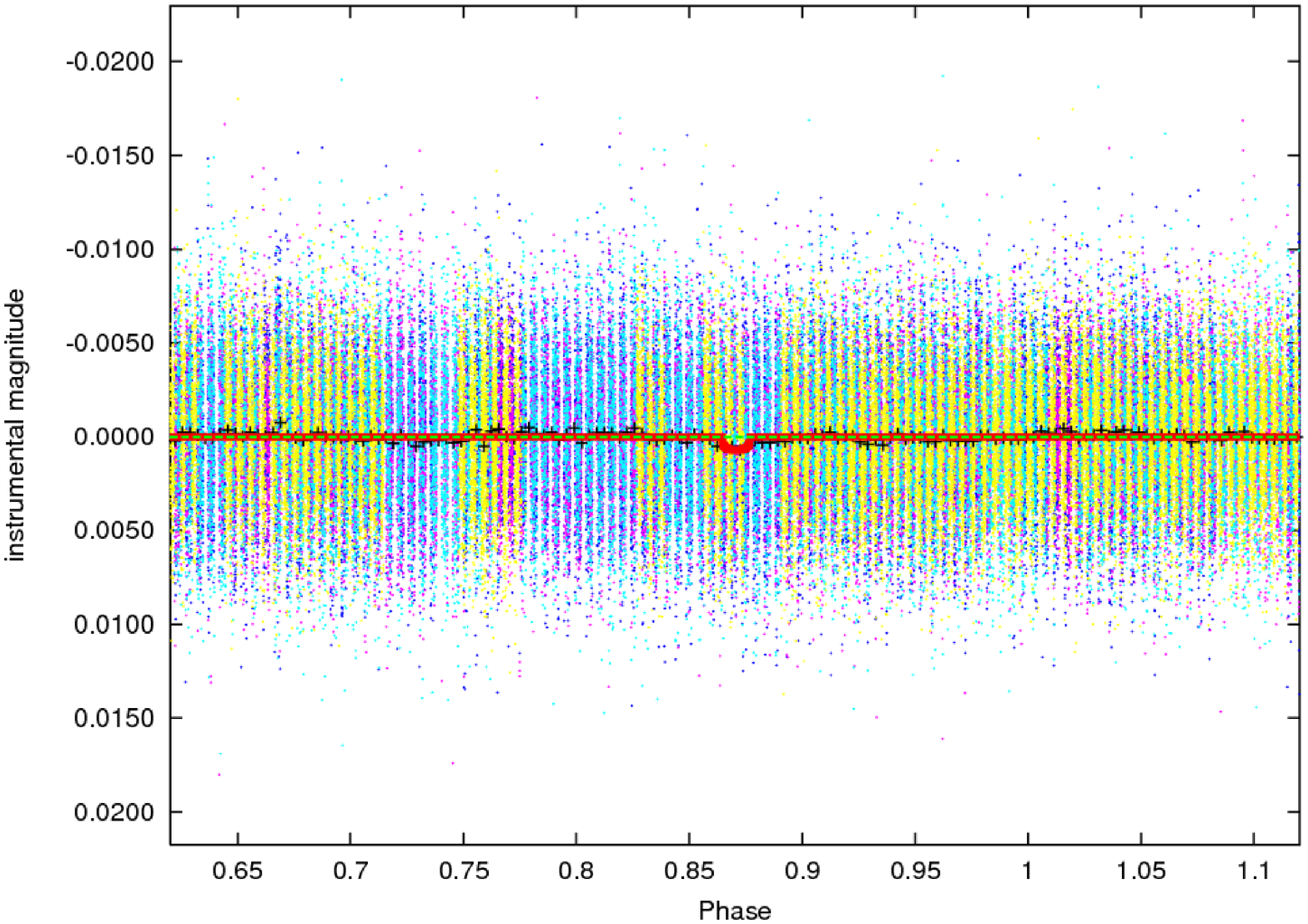}
\includegraphics[width=2.3in, height = 6.0in, angle=270]{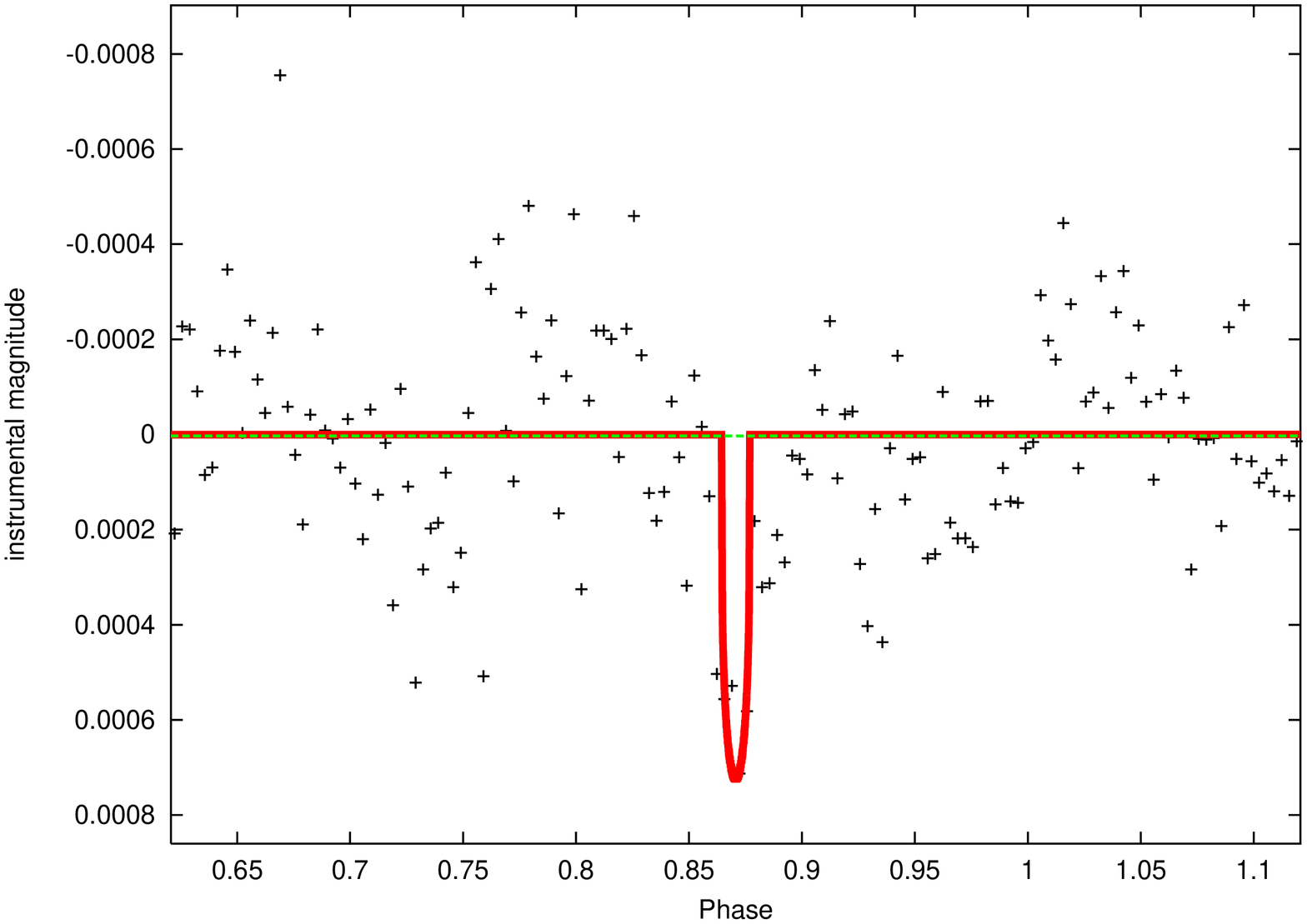}
\includegraphics[width=2.3in, height = 6.0in, angle=270]{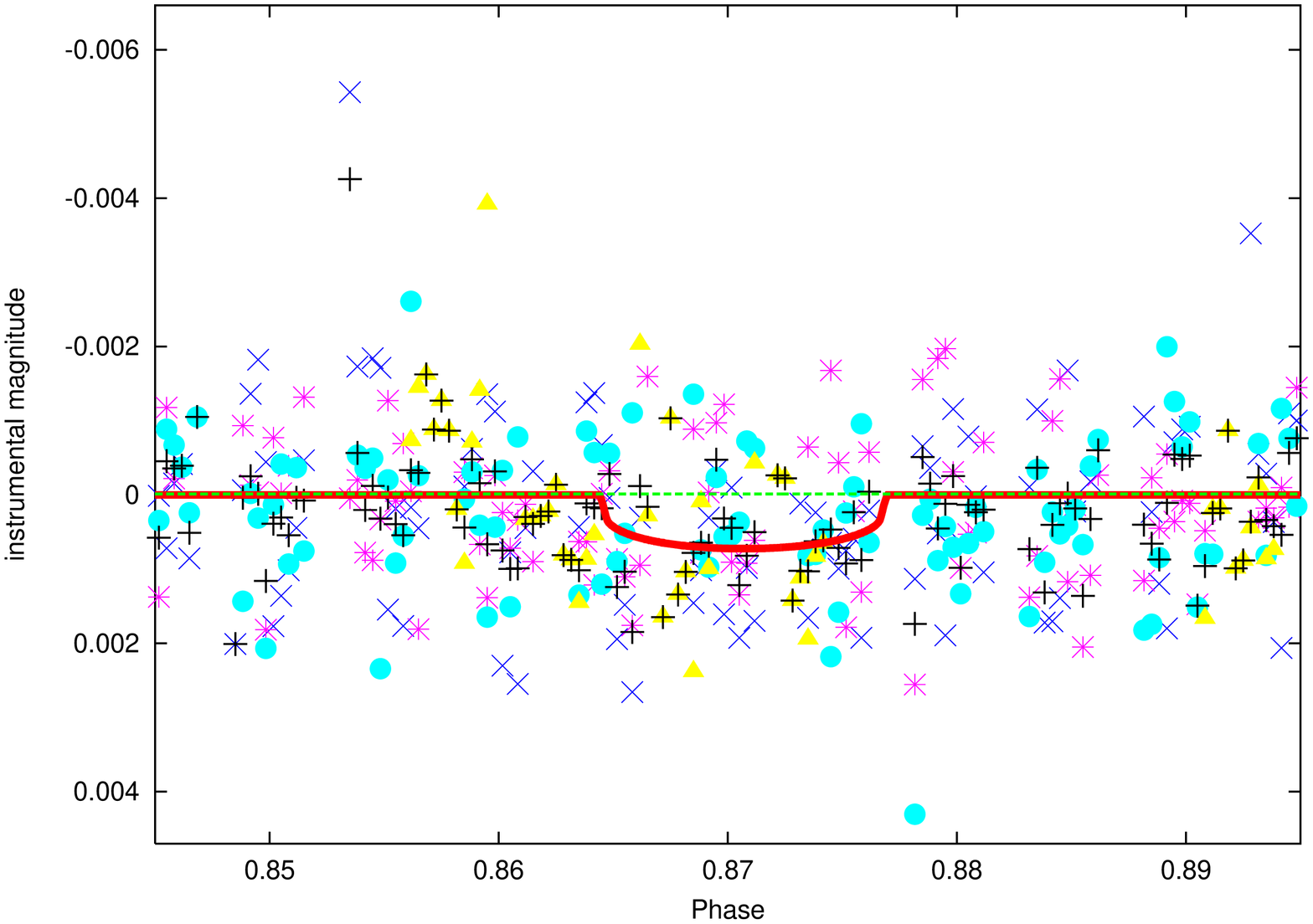}
\caption{
	{The best transit candidate as identified by our analysis of {\it MOST}'s HD 209458 2004
	and 2005 data-sets, but with marginal significance. The thick red-line represents the transit model,
	while the thin green-dashed line represents the constant
	brightness model.
	At top the data unbinned in 2004 (yellow) and the first (cyan), second (magneta), and third (blue) putative transit in 2005, and binned (black).
	Middle: the binned data only, at a different vertical scale.
	At bottom, an expanded portion of the phase diagram, with the binned data from the various
	transits given by the same colours as above. 
	As this candidate failed the improvement in transit over anti-transit criterion, $\Delta\chi^{2}\%$/$\Delta\chi^{2}_{-}\%$ $\ge$ \TransAntiBelieve,
	\ we report it as a possible, but unlikely candidate.
	The period and radius of this putative planet would be approximately
	14.3$d$ and 0.26 $R_{J}$ (2.9$R_{\oplus}$).}  
	\label{FigCandidate}
}

\end{figure}

The {\it MOST} HD 209458 2004 and 2005 data-sets were submitted to the analysis outlined in $\S$\ref{SecTrans}.
There were no transiting planet candidates that met the detection criteria as outlined in $\S$\ref{SecTransDetect}. 
The details of the candidate with the greatest improvement in $\chi^{2}$ are given in Table \ref{TableCandidate}, and
are shown in Figure \ref{FigCandidate}. As this transiting candidate did not meet the detection criterion, 
($\Delta\chi^{2}\%$/$\Delta\chi^{2}_{-}\%$ $\approx$ \PlanetChiPerc \ $<$ $\TransAntiBelieve$), as outlined
in $\S$\ref{SecTransDetect} we do not report it as a putative transiting candidate. 
This event, with a period of, $P \approx \PlanetPeriod$$d$, 
has modest statistical significance, and thus 
we report it is a possible, but unlikely, candidate. The characteristics of this planet are
P = \PlanetPeriod $d$, $R_{p}$ = \PlanetRadius $R_{J}$ (\PlanetRadiusEarth $R_{\oplus}$), 
and $i$ = \PlanetInc$^{o}$.
If this planet is a super-Earth ($\rho$ $\approx$ 3000 kg m$^{-3}$) the mass of this putative planet would be
approximately \PlanetMassEarth \ $M_{\earth}$ (\PlanetMassJupiter \ $M_{J}$).
A transit time combined with the supposed period is given in Table \ref{TableCandidate}. 

Although this event is at a period approximately 4-times that of the known planet, it is not expected that this event
is related to an alias of the data due to the gaps associated
with removing the transits of the known planet. The EEBLS algorithm is dissimilar to a Fourier transform in this regard, 
as gapped data of a certain
period do not induce significant aliases at that period.   
Thus removing the transits
of the known planet should have fully removed all signal at the period, harmonics, and subharmonics of the known planet.
This has been confirmed by limited numerical tests
using data with the same time-sampling as the data used in this application.
A significant signal is not observed 
at harmonics, sub-harmonics of the known planet's orbital period statistically often in these tests.
This putative signal at $P \approx \PlanetPeriod$$d$ is therefore intrinsic to the data.
Evidence for a transit at this period is marginal, but additional {\it MOST} photometry of the HD 209458
system should confirm or disprove its existence.

The most significant brightening event was one observed with a period of approximately $P \approx 8.96 d$.
It is likely statistical in nature, and thus is not expected to be related to any specific astrophysical process.

As an additional sanity check this transit search method was applied to the current data-set
without the removal of the transits
of the known planet, HD 209458b. As expected, the routine correctly uncovers the transit of HD 209458b to a high
degree of accuracy.
The actual parameters of HD 209458b \citep{Knutson} were recovered within
0.0004\% and 0.4\% in period and phase, respectively, while the inclination angle
and radius were recovered within 1.5$^{o}$, and 0.02$R_{J}$ of the actual parameters.  

\begin{deluxetable}{cc}
\tabletypesize{\footnotesize}
\tablecaption{Transit candidate parameters \label{TableCandidate}}
\tablewidth{0pt}
\tablehead{\colhead{parameter}&		\colhead{value} \\}
\startdata

$P$								& \PlanetPeriodLong d \\
$R_{p}$ 							& \PlanetRadius $R_{J}$ (\PlanetRadiusEarth $R_{\oplus}$) \\ 
$i$								& \PlanetInc$^{o}$ \\
$\phi$								& \PlanetPhi \\
Ephemeris Minimum (JD-2451545)					& \PlanetEphemeris \\
$\Delta\chi^{2}\%$/$\Delta\chi^{2}_{-}\%$ 			& \PlanetChiPerc \\
Mass (assuming $\rho$ $\approx$ 3000 kg m$^{-3}$)		& \PlanetMassJupiter \ $M_{J}$ (\PlanetMassEarth \ $M_{\earth}$)\\
\enddata
\end{deluxetable}

\section{Discussion}
\label{SecDiscuss}

{\it MOST}'s 2004 and 2005 observations of HD 209458
have been searched for evidence of other exoplanets in the system. 
A transit-search routine has been adapted to search for
more realistic limb-darkened transits to take advantage of the precise, near-continuous photometry
returned by {\it MOST} and the fact that the 
stellar characteristics of the star, HD 209458, are well established.
Monte Carlo statistics were generated using this routine and indicate that
this routine in combination with the aforementioned {\it MOST} data has placed
unparalleled limits on the size of transiting bodies that have been ruled out in this system 
for a range of periods, and inclination angles. In the most optimistic case 
of edge-on transits planets with radii greater than
\Rjupsmall \ $R_{J}$ ($\sim$\Rearthsmall \ $R_{\oplus}$) to \Rjuplarge \ $R_{J}$ ($\sim$\Rearthlarge \ $R_{\oplus}$) with
periods from half-a-day to two weeks, respectively,
have been ruled out with 95\% confidence through this analysis. Specifically, we have been able to rule out
transiting planets in this system with radii greater
than those given in Figure \ref{FigMonte}. 
This work has constrained theories that invoke smaller exoplanets in the system to explain 
the anomalously large size of HD 209458b, and those that predict ``hot Earths'' due to the inward migration of 
HD 209458b.

\subsection{Mass constraints on other Exoplanets in the system}
\label{SecMass}

The above limits on the smallest planetary radii excluded by {\it MOST}
for transiting planets allow some estimates on the type of planets
that are being excluded. Planet radii change with time as the planet
cools, so we make use of the fact that the age of the HD 209458 system
is determined to be about 5 Gyr with an uncertainty of 1.0 to 1.5 Gyr
\citep{Cody}. This means that we can safely refer to cooling
models similar to the planets in the solar system. One exception would
be the consideration of possible planets in orbits smaller than that of
HD209458b, where tidal heating will lead to larger radii per given mass
and composition.

Our radius limit of 2.2-4.0 $R_{\oplus}$ already places us in the realm of super-Earth
and hot Neptune planets. The range of bulk compositions could encompass
iron-rich cores (super-Mercuries, $\sim 80$\% $Fe$ or $Fe_{0.8}(FeS)_{0.2}$
core), super-Earths (less than 50\% $Fe$ core), water-rich super-Earths
(more than $\sim 10$\% $H_2O$), and water giants (hot Neptunes). Internal
structure models for such planets have been published by \citeauthor{ValenciaA} (\citeyear{ValenciaA}, \citeyear{ValenciaB}).
Hot Neptunes will have structures similar to
Neptune (17.2 $M_{\oplus}$, 3.9 $R_{\oplus}$) and Uranus (14.4 $M_{\oplus}$, 4.0 $R_{\oplus}$).
Planets with
significant amount of water should be referred to as `ice planets', because
water ices VII \& X would form at high pressures of a few GPa. At partial
pressures exceeding $\sim 100$MPa ice planets are unlikely to evaporate
(in 5 Gyr) in orbits larger than 0.1~AU \citep{ValenciaB}, while a pure H+He planet might \citep{Yelle}.

In summary, our radius limit cannot exclude super-Mercuries of any $Fe$ content,
as their masses would have to exceed that of HD209458b itself. It can exclude
some super-Earths, with masses larger than $\sim 8~M_{\oplus}$. It excludes water-rich
super-Earths with masses larger than $\sim 6~M_{\oplus}$ for 50\% water content, and
$\sim 5.5~M_{\oplus}$ for ice planets. The latter two types of planets will have radii
that are very sensitive to stellar heating (hence their orbit) and tidal heating.

HD209458 is a system that has experienced migration and recent simulations
for such planetary systems by
\citet{Raymond} have confirmed previous expectations that
water-rich Earths and super-Earths could often end up in orbits of 0.05 to 0.3~AU.
Our transit search with {\it MOST} rules out super-Earths and hot Neptunes
from 0.01 to 0.12~AU for HD209458 
for edge-on inclinations. 
Planets with inclination angles substantially different then edge-on (see
Figure \ref{FigMonte}) cannot be ruled out with this photometry, except for very short
periods
The radius limits presented here,
combined with the mass-limits from transit-timing analyses (\citealt{Miller}; \citealt{Agol}),
places very firm constraints
on the size and masses of bodies that could still reside in close orbits to HD 209458.
Reobservations of this system with {\it MOST}
should be able to further reduce the size of planets that could remain undetected in this
system, and further constrain theories invoking these planets in nearby orbits to HD 209458b. 
It will be very interesting to use {\it MOST} and obtain similar results
for HD189733 and HAT-P-1 and constrain such theoretical work.

\subsection{Radius and Mass constraints on Trojans}
We can also place a limit on the size of Trojans that may be leading or trailing HD 209458b, assuming the Trojan consistently transits
HD 209458 with an orbital inclination close to that of the known planet (87$^{o}$).
We use the definition of Trojans given by \citet{Ford}
of objects occupying the L4 or L5 stable Lagrangian points.
Trojans lagging or leading HD 209458b with a radius above $R$ = 0.30 $R_{J}$ (3.4 $R_{\oplus}$) should
have been detected with this analysis with 95\% confidence.
The 99\% confidence contour is only slightly larger, $R$ = 0.31 $R_{J}$ (3.5 $R_{\oplus}$).
If these putative planets
were super-Earth analogues ($\rho$ $\approx$ 3000 kg m$^{-3}$)
the corresponding mass limits would be $M = 21 M_{\oplus}$ and $M = 23 M_{\oplus} $ at 95\% and 99\% confidence, respectively,
while for hot Neptune analogues ($\rho$ $\approx$ 1600 kg m$^{-3}$) the mass limits would be $M = 11 M_{\oplus}$ and $M = 12 M_{\oplus} $.
Depending on the structure of planet that is assumed, these mass limits are comparable
or moderately worse than the 13 $M_{\oplus}$ 99.9\% confidence limit of \citet{Ford} for 
HD 209458. However a detailed study using this {\it MOST} photometry of HD 209458 should be able to drastically improve this limit.
It should be noted that 
\citet{Ford} cast doubt on the hypothesis that a hypothetical Trojan will consistently transit HD 209458. They 
note that if the hypothetical Trojan has a vertical libration amplitude greater than approximately 9$^{o}$ it will not consistently transit HD 209458.

\acknowledgements
We thank the referee, Ron Gilliland, for useful comments that have improved this paper.
We thank Sara Seager for useful discussions on exoplanetary mass models. 
The Natual Sciences and Engineering Research Council of Canada supports the research of D.B.G., J.M.M., A.F.J.M., J.F.R., S.M.R., G.A.H.W., and B.G..
Additional support for A.F.J.M. comes from FCAR
(Qu\'ebec). R.K. and A.W., are supported by the Canadian Space Agency. W.W.W. is supported by the Austrian Space Agency and the Austrian Science Fund (P17580).
The Canadian Foundation for Innovation provided funding for the
LeVerrier Beowulf cluster.

\begin{thebibliography}{}


\bibitem[Agol \& Steffen(2007)]{Agol}
Agol, E. \& Steffen, J.H. 2007, \mnras, 374, 941

\bibitem[Alonso et al.(2004)]{Alonso}
Alonso, R., Brown, T.~M., Torres, G., Latham, D.~W., Sozzetti, A., Mandushev, G., Belmonte, J.~A., Charbonneau, D., Deeg, H.~J., 
Dunham, E.~W., O'Donovan, F.~T., Stefanik, R.~P. 2004, \apjl, 613, L153

\bibitem[Alonso et al.(2004)]{AlonsoB}
Alonso, R., Deeg, H. J., Brown, T. M., Belmonte, J. A. 2004, Astronomische Nachrichten, 325, 594-597.


\bibitem[Baglin (2003)]{Baglin}
Baglin, A. 2003, Advances in Space Research, Volume 31, 345-349.

\bibitem[Bakos et al.(2002)]{Bakos}
Bakos, G.~{\'A}, L{\'a}z{\'a}r, J., Papp, I., S{\'a}ri, P., Green, E.~M. 2002, \pasp, 114, 974

\bibitem[Bakos et al.(2006)]{BakosB}
Bakos, G.~A., Noyes, R.~W., Kovacs, G., Latham, D.~W., Sasselov, D.~D., Torres, G., Fischer, D.~A., Stefanik, R.~P.,
Sato, B., Johnson, J.~A., Pal, A., Marcy, G.~W., Butler, R.~P., Esquerdo, G.~A., Stanek, K.~Z., Lazar, J., Papp, I., Sari, P., Sipocz, B.
2007, 656, 552

\bibitem[Barge et al.(2005)]{Barge}
Barge, P., Baglin, A., Auvergne, M., Buey, J.-T., Catala, C., Michel, E., Weiss, W. W., Deleuil, M., Jorda, L., Moutou, C.,
COROT Team 2005, SF2A-2005: Semaine de l'Astrophysique Francaise, eds. F. Casoli, T. Contini, J.M. Hameury \& L. Pagani, EdP-Sciences, Conference Series, p. 193.

% Kepler Paper Review Article Good
\bibitem[Basri et al.(2005)]{Basri}
Basri, G., Borucki, W.~J., Koch, D. 2005, New Astronomy Review, 49, 478.

\bibitem[Bodenheimer et al.(2001)]{Bodenheimer}
Bodenheimer, P., Lin, D.~N.~C., Mardling, R.~A. 2001, \apj, 548,466

\bibitem[Bodenheimer et al.(2003)]{BodenheimerThree}
Bodenheimer, P., Laughlin, G., Lin, D.~N.~C. 2003, \apj, 592,555

% Former Corot Paper
\bibitem[Bord{\'e} et al.(2001)]{Borde}
Bord{\'e}, P., Rouan, D., L{\'e}ger, A. 2001,  Academie des Sciences Paris Comptes Rendus Serie Physique Astrophysique, 7, 1049.


\bibitem[Borucki et al.(2004)]{Borucki}
Borucki, W.; Koch, D.; Boss, A.; Dunham, E.; Dupree, A.; Geary, J.;
Gilliland, R.; Howell, S.; Jenkins, J.; Kondo, Y.; Latham, D.; Lissauer,
J.; Reitsema, H. 2004, Second Eddington Workshop: Stellar structure and
habitable planet finding, eds. F. Favata, S. Aigrain \& A. Wilson,
ESA SP-538, Noordwijk: ESA Publications Division, 177 - 182.

\bibitem[Burke et al.(2006)]{Burke}
Burke, C.J., Gaudi, B.S., DePoy, D.L., Pogge, R.W. 2006, \aj, 132, 210

\bibitem[Charbonneau et al.(2000)]{Charbonneau}
Charbonneau, D., Brown, T.M., Latham, D.W., Mayor, M. 2000, \apj, 529,L45

\bibitem[Cody \& Sasselov(2002)]{Cody}
Cody, A.M. \& Sasselov, D. 2002 ApJ, 569, 451

\bibitem[Ford \& Gaudi (2006)]{Ford}
Ford, E., Gaudi, B.S. 2006, \apjl, submitted, astro-ph/0609298

\bibitem[Gilliland et al.(2000)]{Gilliland}
Gilliland, R.~L., Brown, T.~M., Guhathakurta, P., Sarajedini, A., Milone, E.~F., Albrow, M.~D., Baliber, N.~R., Bruntt, H., Burrows, A.,
Charbonneau, D., Choi, P., Cochran, W.~D., Edmonds, P.~D., Frandsen, S., Howell, J.~H., Lin, D.~N.~C., Marcy, G.~W.,  Mayor, M.,
Naef, D., Sigurdsson, S., Stagg, C.~R., Vandenberg, D.~A., Vogt, S.~S., Williams, M.~D. 2000, \apjl, 545, L47

\bibitem[Henry et al.(2000)]{Henry}
Henry, G.~W., Marcy, G.~W., Butler, R.~P., Vogt, S.~S. 2000, \apjl, 529, L41

\bibitem[Hidas et al.(2005)]{Hidas}
Hidas, M.~G.,  Ashley, M.~C.~B.,  Webb, J.~K., Irwin, M.,  Phillips, A.,  Toyozumi, H., Derekas, A., 
Christiansen, J.~L.,  Nutto, C.,  Crothers, S. 2005, \mnras, 360, 703

\bibitem[Hood et al.(2005)]{Hood}
Hood, B., Cameron, A.~C., Kane, S.~R.,  Bramich, D.~M., Horne, K., Street, R.~A., Bond, I.~A.,  Penny, A.~J.,
Tsapras, Y., Quirrenbach, A., Safizadeh, N., Mitchell, D., Cooke, J. 2005, \mnras, 360, 791

\bibitem[Ida \& Lin(2004)]{Ida}
Ida, S., Lin, D.~N.~C. 2005, \apj, 626, 1045.

\bibitem[Knutson et al.(2007)]{Knutson}
Knutson, H., Charbonneau, D., Noyes, R.W., Brown, T.M., Gilliland, R.L. 2007, \apj, 655, 564

\bibitem[Kov{\'a}cs (2003)]{KovacsWEB}
Kov{\'a}cs, G. 2003, http://www.konkoly.hu/staff/kovacs/index.html

\bibitem[Kov{\'a}cs et al.(2002)]{Kovacs}
Kov{\'a}cs, G., Zucker, S., Mazeh, T. 2002, \aap, 391, 369

\bibitem[Laughlin et al.(2005)]{Laughlin}
Laughlin, G., Marcy, G.~W., Vogt, S.~S., Fischer, D.~A. 2005, \apjl, 629, L121

\bibitem[Levenberg(1944)]{Leven} Levenberg, K. 1944, Quart. Appl. Math. 2, 164

\bibitem[Mandel \& Agol(2002)]{Mandel}
Mandel, K., Agol, E. 2002, \apj, 580, L171

\bibitem[Marquardt(1963)]{Marq} Marquardt, D.W. 1963, SIAM J. Appl. Math, 11, 431

\bibitem[Matthews et al.(2004)]{Matthews}
Matthews, J.M., Kusching, R., Guenther, D.B., Walker, G.A.H., Moffat, A.F.J., Rucinski, S.M., Sasselov, D., Weiss, W.W. 2004, Nature, 430, 51

\bibitem[Mayor \& Queloz(1995)]{Mayor}
Mayor, M., Queloz, D. 1995, \nat, 378, 355

\bibitem[Mazeh et al.(2000)]{Mazeh}
Mazeh, T., Naef, D., Torres, G., Latham, D.~W., Mayor, M., Beuzit, J.-L., Brown, T.~M., Buchhave, L., Burnet, M., Carney, B.~W., 
Charbonneau, D., Drukier, G.~A., Laird, J.~B., Pepe, F., Perrier, C., Queloz, D., Santos, N.~C., Sivan, J.-P., Udry, S., Zucker, S. 
2000, \apjl, 532, L55

\bibitem[McCullough et al.(2006)]{McCullough}
McCullough, P.~R., Stys, J.~E., Valenti, J.~A., Johns-Krull, C.~M.,  Janes, K.~A., Heasley, J.~N., 
Bye, B.~A., Dodd, C., Fleming, S.~W., Pinnick, A., Bissinger, R., Gary, B.~L., Howell, P.~J., Vanmunster, T.
2006, \apj, 648, 1228

\bibitem[Miller-Ricci et al.(2006)]{Miller}
Miller-Ricci, E., Rowe, J.F., Sasselov, D., Matthews, J.M., Guenther, D.B., Kuschnig, R., Moffat, A.F.J., Rucinski, S.M., Walker, G.A.H., Weiss, W.W.
2006, \apj, in preparation

\bibitem[Mochejska et al.(2006)]{Mochejska}
Mochejska, B.~J., Stanek, K.~Z.,  Sasselov, D.~D., Szentgyorgyi, A.~H., Adams, E.,  Cooper, R.~L., Foster, J.~B., Hartman, J.~D., Hickox, R.~C., Lai, K., Westover, M., Winn, J.~N.
2006, \aj, 131, 1090

\bibitem[Press et al.(1992)]{Numer}
Press, W.H., Flannery, B., Teukolsky, S.A., Vetterling, W.T. 1992. Numerical Recipes in Fortran, 2nd ed. (Cambridge University Press, Cambridge, England).

\bibitem[O'Donovan et al.(2006)]{ODonovan}
O'Donovan, F.~T., Charbonneau, D., Mandushev, G., Dunham, E.~W., Latham, D.~W., Torres, G., Sozzetti, A., 
Brown, T.~M., Trauger, J.~T., Belmonte, J.~A., Rabus, M., Almenara, J.~M., Alonso, R., Deeg, H.~J., Esquerdo, G.~A., Falco, E.~E.,
Hillenbrand, L.~A., Roussanova, A., Stefanik, R.~P., Winn, J.~N. 2006, astro-ph/0609335

\bibitem[Raymond et al.(2006)]{Raymond}
Raymond, S., Mandell, A., \& Sigurdsson, S. 2006, Science, 313, 1413

\bibitem[Rowe et al.(2006a)]{Rowe}
Rowe, J.F., Matthews, J.M., Seager, S., Kuschnig, R., Guenther, D.B., Moffat, A.F.J., Rucinski, S.M., Sasselov, D., Walker, G.A.H., Weiss, W.W. 2006, \apj, 646, 1241.

\bibitem[Rowe et al.(2007)]{Rowe2}
Rowe, J.F., Matthews, J.M., Seager, S., Kuschnig, R., Guenther, D.B., Moffat, A.F.J., Rucinski, S.M., Sasselov, D., Walker, G.A.H., Weiss, W.W. 2007, \apj, in preparation.

\bibitem[Sahu et al.(2006)]{Sahu}
Sahu, K.C., et al. 2006, Nature, 443, 534

\bibitem[Schneider (2006)]{Schneider}
Schneider, J. 2006, http://exoplanet.eu/

\bibitem[Steffen \& Agol (2005)]{Steffen}
Steffen, J.H., Agol, E. 2005, \mnras, 364, L96

\bibitem[Street et al. (2004)]{Street}
Street, R.~A., Christian, D.~J., Clarkson, W.~I., Collier Cameron, A., Evans, N., Fitzsimmons, A., 
Haswell, C.~A., Hellier, C., Hodgkin, S.~T., Horne, K., Kane, S.~R., Keenan, F.~P., Lister, T.~A., Norton, A.~J., 
Pollacco, D., Ryans, R., Skillen, I., West, R.~G., Wheatley, P.~J. 2004, Astronomische Nachrichten, 325, 565

\bibitem[Thommes (2005)]{Thommes}
Thommes, E.W. 2005, \apj, 626, 1033

\bibitem[Tingley (2003)]{Tingley}
Tingley, B. 2003, \aap, 408, L5

\bibitem[Udalski et al.(2002)]{Udalski}
Udalski, A., Paczynski, B., Zebrun, K., Szymaski, M., Kubiak, M., Soszynski, I., Szewczyk, O., Wyrzykowski, L., Pietrzynski, G.
2002, Acta Astronomica, 52, 1

\bibitem[Valencia et al.(2006)]{ValenciaA}
Valencia, D., O'Connell, R., \& Sasselov, D. 2006, Icarus, 181, 545

\bibitem[Valencia et al.(2007)]{ValenciaB}
Valencia, D., Sasselov, D., \& O'Connell, R. 2007, \apj, 656, 545

\bibitem[von Braun et al.(2005)]{vonBraun}
von Braun, K., Lee, B.~L., Seager, S., Yee, H.~K.~C., 
Mall{\'e}n-Ornelas, G.,  Gladders, M.~D. 2005, \pasp, 117, 141

\bibitem[Walker et al.(2003)]{Walker}
Walker, G. A. H., et al. 2003, \pasp, 115, 1023

\bibitem[Weldrake et al.(2005)]{Weldrake}
Weldrake, D.~T.~F., Sackett, P.~D., Bridges, T.~J., Freeman, K.~C. 2005, \apj, 620, 1043

\bibitem[Wittenmyer et al.(2005)]{Wittenmyer}
Wittenmyer, R.A., Welsh, W.F., Orosz, J.A., Schultz, A.B., Kinzel, W., Kochte, M., Bruhweiler, F., Bennum, D., Henry, G.W., Marcy, G.~W.,  Fischer, D.~A., Butler, R.~P., Vogt, S.~S. 2005, \apj, 632, 1157

\bibitem[Yelle(2004)]{Yelle}
Yelle, R. 2004, Icarus, 170, 167

\bibitem[Zhou et al.(2005)]{Zhou}
Zhou, J.L., Aarseth, S.J., Lin, D.N.C., Nagasawa, M. 2005, \apjl, 631, L85

\end{thebibliography}
\end{document}